\documentclass[fleqn,usenatbib]{mnras}
\usepackage{newtxtext,newtxmath}
\usepackage[T1]{fontenc}
\usepackage{aecompl}
\DeclareRobustCommand{\VAN}[3]{#2}
\let\VANthebibliography\thebibliography
\def\thebibliography{\DeclareRobustCommand{\VAN}[3]{##3}\VANthebibliography}
\usepackage{graphicx}
\usepackage{amsmath}
\usepackage{caption}
\usepackage{subcaption}

\usepackage{amssymb}
\usepackage{pdflscape}
\usepackage{siunitx}
\usepackage[symbol]{footmisc}
\title{The origin of the soft X-ray excess in the narrow-line Seyfert 1 galaxy SBS 1353+564}
\author[Xinpeng Xu et al.]{
Xinpeng Xu,$^{1,2}$
Nan Ding,$^{3}$\thanks{Corresponding authors: \href{mailto:orient.dn@foxmail.com}{orient.dn@foxmail.com}; \href{mailto:qsgu@nju.edu.cn}{qsgu@nju.edu.cn}}
Qiusheng Gu,$^{1,2}$\footnote[1]{}
Xiaotong Guo,$^{1,2}$
E. Contini$^{1,2}$
\\
$^{1}$School of Astronomy and Space Science, Nanjing University, Nanjing, Jiangsu 210093, People's Republic of China\\
$^{2}$Key Laboratory of Modern Astronomy and Astrophysics (Nanjing University), Ministry of Education, Nanjing 210093, People's Republic of China\\
$^{3}$School of Physical Science and Technology, Kunming University, Kunming 650214, People's Republic of China\\
}
\date{Accepted x. Received y; in original form z}
\pubyear{2021}
\begin{document}
\label{firstpage}
\pagerange{\pageref{firstpage}--\pageref{lastpage}}
\maketitle
\begin{abstract}

We present for the first time the timing and spectral analyses for a narrow-line Seyfert 1 galaxy, SBS 1353+564, using \textit{XMM-Newton} and \textit{Swift} multi-band observations from 2007 to 2019. Our main results are as follows: 1) The temporal variability of SBS 1353+564 is random, while the hardness ratio is relatively constant over a time span of 13 years; 2) We find a prominent soft X-ray excess feature below 2 keV, which cannot be well described by a simple blackbody component; 3) After comparing the two most prevailing models for interpreting the origin of the soft X-ray excess, we find that the relativistically smeared reflection model is unable to fit the data above 5 keV well and the X-ray spectra do not show any reflection features, such as the Fe K${\alpha}$ emission line. However, the warm corona model can obtain a good fitting result. For the warm corona model, we try to use three different sets of spin values to fit the data and derive different best-fitting parameter sets; 4) We compare the UV/optical spectral data with the extrapolated values of the warm corona model to determine which spin value is more appropriate for this source, and we find that the warm corona model with non-spin can sufficiently account for the soft X-ray excess in SBS 1353+564. 

\end{abstract}


\section{Introduction}

Narrow-line Seyfert 1 galaxies (NLS1s) are a unique type of Active Galactic Nuclei (AGNs), which are powered by the accretion of supermassive black holes. NLS1s are characterized and defined by the following features: weak [O {\sevensize III}] emission with the flux ratio [O {\sevensize III}]${\lambda}$5007/H${\beta}$ < 3, Balmer lines with the Full Width at Half Maximum ($\text{FWHM}_{H\beta}$) < 2000 km $\text{s}^{-1}$, and strong Fe {\sevensize II} emission lines \citep{1985ApJ...297..166O, 1989ApJ...342..224G}. In addition, NLS1s are believed to have smaller black hole masses compared to the broad-line Seyfert 1 galaxies (BLS1s) with the same luminosity, usually ranging from ${10}^{6}$ $\text{M}_{\odot}$ to ${10}^{8}$ $\text{M}_{\odot}$ \citep{2004ApJ...606L..41G}. It is also suggested that NLS1s are accreting at a very high rate, close to or even beyond the Eddington limit \citep{2004AJ....127.1799G, 2004ApJ...606L..41G}. The X-ray spectra of NLS1s generally consist of a primary X-ray continuum in the form of a power law \citep{1980A&A....86..121S}, a prominent soft X-ray excess below $\sim$ 2 keV \citep{1985MNRAS.217..105A, 1996A&A...305...53B}, Fe K${\alpha}$ emission lines at $\sim$ 6.4 keV \citep{2007MNRAS.382..194N}, and sometimes a Compton reflection hump above 20 keV \citep{1991MNRAS.249..352G}. 

The origin of the soft X-ray excess is widely disputed. Nowadays, the most prevailing interpretations are the relativistically smeared reflection from the ionized accretion disk \citep{2005MNRAS.358..211R, 2006MNRAS.365.1067C, 2013MNRAS.428.2901W} and the thermal Comptonisation from a warm corona \citep{1998MNRAS.301..179M, 2004MNRAS.349L...7G, 2012MNRAS.420.1848D}. The relativistically smeared reflection model interprets the soft excess as the consequence of the hard X-ray photons reflected from the surface of the disk. The seed UV/optical photons produced in the disk are Compton up-scattered to the hard X-ray band in a hot optically thin corona above the central black hole. A fraction of the hard X-ray photons are incident to the surface of the accretion disk and are reflected to produce the low-energy emission-line clusters in the soft X-ray band, which are smeared by the strong relativistic effects. These smeared emission lines contribute to the soft X-ray excess that we observe. The relativistically smeared reflection model has been widely used to explain the physical origin of the X-ray emission in many Seyfert 1 galaxies, such as 1H 0323+342 \citep{2020MNRAS.496.2922M}; IRAS 09149-6206 \citep{2020MNRAS.499.1480W}; 1H 0707-495 \citep{2021A&A...647A...6B}.

\begin{table*}
 \caption{The summary of \textit{XMM-Newton} and \textit{Swift} observations of SBS 1353+564. In columns (5) and (6), the net exposures and the net count rates for \textit{XMM-Newton} are derived from the EPIC-pn instrument, and those for \textit{Swift} are derived from the XRT detector. The time intervals dominated by background flares are removed from the net exposures of EPIC-pn. In column (7), we list the exposures in the V / B / U / UVW1 / UVM2 / UVW2 filters for \textit{Swift} UVOT detector, and the symbol `${...}$' denotes no exposure in this particular filter.}
 \label{tab:obs}
 \begin{tabular}{ c c c c c c c }
  \hline
  (1) & (2) & (3) & (4) & (5) & (6) & (7)\\
  Observation & Obs. ID & Start Date & Duration & Exposure & Count Rate & UVOT Exposure\\
    &   & (yyyy-mm-dd) & (s) & (s) & (counts $\text{s}^{-1}$) & (s)\\
  \hline
  \textit{XMM-Newton}& 0741390201 & 2014-06-21 & 25000 & 16460 & ${5.162\pm0.018}$ &\\
    & 0741390401 & 2014-07-01 & 28000 & 18560 & ${2.390\pm0.011}$ &\\
  \hline
  \textit{Swift}& 00036547001 & 2007-05-13 &   & 4442 & ${0.101\pm0.005}$ & 363 / 363 / 363 / 736 / 958 / 1474\\
    & 00036547002 & 2007-05-21 &   & 977 & ${0.239\pm0.016}$ & ... / 158 / 157 / 316 / ... / 313\\
    & 00036547003 & 2007-06-20 &   & 3468 & ${0.090\pm0.005}$ & 301 / 301 / 301 / 601 / 787 / 1204\\
    & 00036547004 & 2007-06-26 &   & 3983 & ${0.136\pm0.006}$ & 329 / 329 / 329 / 659 / 889 / 1322\\
    & 00036547005 & 2007-06-28 &   & 1070 & ${0.131\pm0.011}$ & 65 / 65 / 66 / 408 / 174 / 260\\
    & 00091336001 & 2013-01-06 &   & 1492 & ${0.183\pm0.011}$ & 86 / 157 / 157 / 315 / 193 / 543\\
    & 00091336002 & 2013-02-20 &   & 594 & ${0.076\pm0.012}$ & 42 / 42 / 42 / 85 / 167 / 170\\
    & 00091867001 & 2014-04-11 &   & 1246 & ${0.259\pm0.015}$ & 101 / 101 / 101 / 203 / 310 / 405\\
    & 00091867002 & 2014-04-20 &   & 484 & ${0.145\pm0.017}$ & ... / ... / ... / 480 / ... / ...\\
    & 00091867004 & 2014-09-02 &   & 1059 & ${0.172\pm0.013}$ & ... / ... / ... / ... / 1054 / ...\\
    & 00091867005 & 2014-09-03 &   & 313 & ${0.163\pm0.023}$ & ... / ... / ... / 312 / ... / ...\\
    & 00093084001 & 2017-10-19 &   & 331 & ${0.279\pm0.029}$ & ... / 55 / 55 / 110 / ... / 95\\
    & 00093084003 & 2018-02-23 &   & 276 & ${0.138\pm0.023}$ & ... / ... / ... / 274 / ... / ...\\
    & 00095079001 & 2019-10-12 &   & 720 & ${0.191\pm0.016}$ & 57 / 57 / 57 / 113 / 186 / 226\\
    & 00095079003 & 2019-11-13 &   & 552 & ${0.259\pm0.022}$ & 43 / 43 / 43 / 85 / 144 / 171\\
  \hline  
 \end{tabular}
\end{table*}

The warm corona model is also favored to explain the soft excess. Like the reflection model, it assumes an optically thin (${\tau}$ $\sim$ 1), high temperature ($\sim$ 100 keV) corona which produces the Comptonized primary continuum. Apart from the hot corona, the warm corona model assumes that the soft X-ray photons come from an optically thick (${\tau}$ $\sim$ 10 -- 20), low temperature ($\sim$ 0.1 -- 1 keV) plasma above the surface of the accretion disk \citep{1998MNRAS.301..179M, 2003A&A...412..317C, 2004MNRAS.349L...7G, 2012MNRAS.420.1848D}. The UV/optical photons from the disk are partly released into the warm corona and are inverse Compton scattered to the soft X-ray band contributing to the soft X-ray excess. The X-ray spectra of many AGNs, such as Zw 229.015 and Ton S180, are proved to be well described by the warm corona model in the past works \citep{2019MNRAS.488.4831T, 2020MNRAS.497.2352M}. 

SBS 1353+564 is a narrow-line Seyfert 1 AGN \citep{2006A&A...455..773V} at redshift $\textit{z}$ = 0.1215 \citep{2017ApJS..229...39R}. It has a Galactic absorption column density of $\textit{N}_\text{H} = 7.80\times10^{19}  \text{cm}^{-2}$ \citep{2016A&A...594A.116H}, and also shows evidence of intrinsic UV absorption \citep{2007AJ....134.1061D}. The FWHM of the broad H${\beta}$ line is measured to be 2126 km $\text{s}^{-1}$, which is slightly larger than 2000 km $\text{s}^{-1}$, defined as the upper limit for NLS1s \citep{2017ApJS..229...39R}. The [O {\sevensize III}] emission strength is [O {\sevensize III}]${\lambda}$5007/H${\beta}$ = 1.87, and the [Fe {\sevensize II}] emission is weak with [Fe {\sevensize II}]/H${\beta}$ = 0.85 \citep{2017ApJS..229...39R}. 

SBS 1353+564 has been classified as a bright soft X-ray selected AGN by ROSAT All-Sky Survey \citep{1998A&A...330...25G, 1999A&A...350..805G}. The source is also considered as a bright object with a very high accretion rate at quasar luminosity, and it is believed to have a very steep hard X-ray photon index with ${\Gamma}_\text{2 -- 10 keV}$ > 2 \citep{2020MNRAS.495.1158C, 2020ApJ...904..200Y}, which agrees with the relation between the Eddington ratio (${\lambda}_\text{Edd}$) and ${\Gamma}_\text{2 -- 10 keV}$ \citep{1999ApJ...526L...5L, 2013MNRAS.433.2485B}. SBS 1353+564 is also classified as a radio-quiet NLS1\citep{2018A&A...614A..87B}, which is consistent with the inverse correlation by \citet{2020ApJ...904..200Y} that super-Eddington AGNs are mostly radio-quiet. The source shows a very prominent soft X-ray excess and a very soft continuum with ${\Gamma}_\text{2 -- 10 keV}$ > 2 in its spectra. In the previous work, \citet{2020MNRAS.498.3888J} has modelled the X-ray spectra of SBS 1353+564 with a reflection-based model and indicates that the source as an ultra-soft NLS1 has a low disk density and a high disk ionization, which makes the emission features in the spectra very weak, contributing to a smooth and prominent soft X-ray excess and a weak iron line feature. However, the weak iron line feature makes it hard to distinguish the reflection model from other models, such as the warm corona model \citep{2020MNRAS.498.3888J}. So it is worth comparing the reflection model and the warm corona model in the spectral analysis of such a soft and smooth NLS1.

SBS 1353+564 was observed two times by \textit{XMM-Newton} and 15 times by \textit{Swift} \citep{2010ApJS..187...64G} with high quality X-ray spectral data. This source has rather high count rates and enough exposure times for both observations by \textit{XMM-Newton}. The two \textit{XMM-Newton} observations Obs. ID 0741390201 and Obs. ID 0741390401 were performed with only ten days apart. There is practically no variability in spectral shape between the two observations, however, the flux decreases by a factor of $\sim$ 2 in merely ten days. Such a significant flux drop in such a short time is quite interesting for a narrow-line Seyfert 1 AGN. Besides, the multiple high-quality XMM-Newton observations are actually quite few for NLS1s, especially for the sources with significant changes in flux in a short timescale. The analysis of the variability in flux may help us to distinguish the models for explaining the soft excess. The \textit{XMM-Newton} and \textit{Swift} data of SBS 1353+564 were processed to derive the light curves and the spectra for the first time. In the timing analysis, we considered both the short- and long-term variability, and we also used the fractional variability ($\textit{F}_\textit{var}$) to probe the temporal variability in SBS 1353+564. In the spectral analysis, we analyzed the multi-epoch X-ray spectra and found a prominent soft X-ray excess feature. We tried both the reflection model and the warm corona model to describe the X-ray spectra. We compared and analyzed the fitting results to find out the possible physical origin of the soft X-ray excess in SBS 1353+564.

This paper is organized as follows. We describe \textit{XMM-Newton} and \textit{Swift} observations and data reduction in Section 2. In Section 3, we examine the variability of the source in both the long- and short-term timescales. In Section 4, we present the spectral analysis of our source. Finally, we discuss our main results and conclusions in Section 5. The standard ${\Lambda}$CDM cosmology with parameters $\textit{H}_\text{0} = 70 \text{ km} \text{ s}^{-1} \text{ Mpc}^{-1}$, ${\Omega}_\text{M} = 0.27$ and ${\Omega}_{\Lambda} = 0.73$ is adopted throughout this work. The luminosity distance of SBS 1353+564 based on this cosmology is 569.21 Mpc.

\begin{figure}
 \includegraphics[width=\columnwidth]{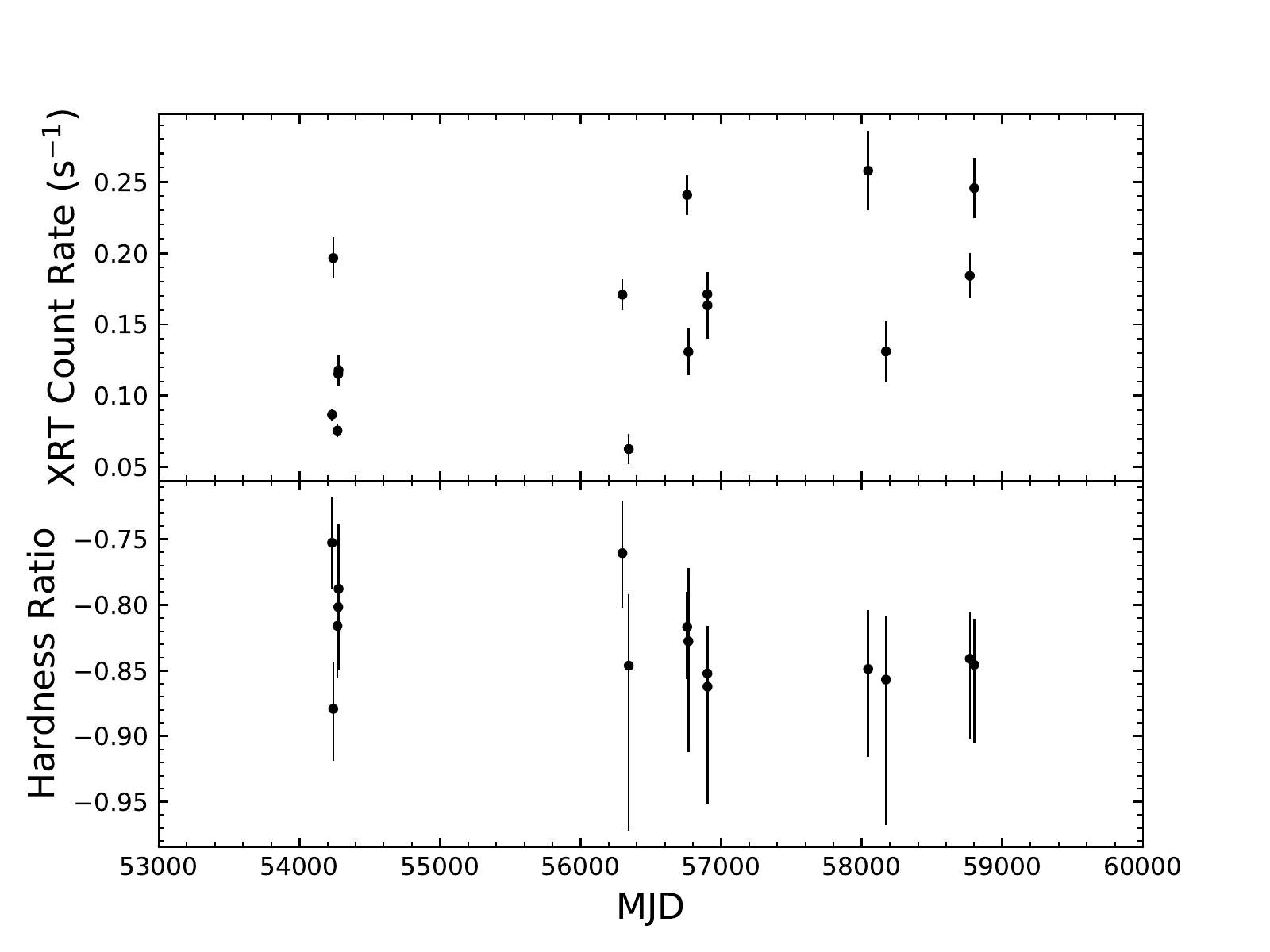}
 \vspace*{-7mm}
 \caption{The long-term \textit{Swift} XRT light curve for SBS 1353+564 in the 0.3 -- 10 keV, together with the hardness ratio depicted below. The x-axis represents the date in MJD (Modified Julian Date). Random variations are detected in the light curve, while the hardness ratio is found to be fairly consistent with a constant.}
 \label{fig:XRT}
\end{figure} 

\section{Observations and Data Reduction}

In this section, we describe \textit{XMM-Newton} and \textit{Swift} \citep{2010ApJS..187...64G} multi-band observations of SBS 1353+564 over 13 years and the procedure of the archival data reduction. We downloaded the \textit{XMM-Newton} and \textit{Swift} data from NASA's High Energy Astrophysics Science Archive Research Center (HEASARC) and processed the data with {\sevensize HEASOFT} v.6.26.1 \citep{2014ascl.soft08004N}. The summary of the observations is listed in Table~\ref{tab:obs}.

\subsection{\textit{XMM-Newton}}

SBS 1353+564 was observed two times with \textit{XMM-Newton} telescope \citep{2001A&A...365L...1J}. The \textit{XMM-Newton} European Photon Imaging Camera (EPIC) instruments were operated in the small window mode with the medium filter applied in both observations. The two observations Obs. ID 0741390201 and Obs. ID 0741390401 were performed in June 21, 2014 and July 1, 2014, exactly ten days apart. The \textit{XMM-Newton} Observation Data File (ODF) were produced with the \textit{XMM-Newton} Science Analysis System ({\sevensize SAS} v.18.0.0 \citep{2004ASPC..314..759G}) and the most recent updated calibration files. We only considered the EPIC-pn detector here, because it has better spectral resolution than the EPIC-MOS detector \citep{2001A&A...365L..18S}. We processed pn data using task \texttt{EPPROC} and got the pn calibrated photon event files. Then we filtered the EPIC-pn data for background flares by creating a Good Time Interval (GTI) file above 10 keV with rate $\leqslant$ 0.4 counts $\text{s}^{-1}$ using the task \texttt{TABGTIGEN}. We also checked for pile-up using the task \texttt{EPATPLOT} and found no presence of pile-up in both the observations. The source spectra were extracted from circular regions with radii of 30$''$ centered on the source, and we extracted the background spectra from nearby off-source circular regions with radii of 70$''$. We selected the single and double events (PATTERN $\leqslant$ 4) and set FLAG == 0 to filter the pn spectra. The {\sevensize SAS} tasks \texttt{RMFGEN} and \texttt{ARFGEN} were used to generate the redistribution matrix files (RMF) and the ancillary response files (ARF), respectively. The extracted spectra were grouped into a minimum of 100 counts per bin using the task \texttt{GRPPHA}. The net exposure times of EPIC-pn instrument are 16.46 ks for Obs. ID 0741390201 and 18.56 ks for Obs. ID 0741390401. As for the variability, we extracted the background subtracted light curves using \texttt{EPICLCCORR} with a time binsize of 500 s. 

\subsection{\textit{Swift}}

\textit{Swift} observed SBS 1353+564 for 15 times in total from May 13, 2007 to November 13, 2019 \citep{2010ApJS..187...64G}. In our work, we analyzed both XRT and UVOT data. We processed the X-ray Telescope (XRT) \citep{2005SSRv..120..165B} data of all the 15 observations using the task \texttt{XRTPIPELINE}. All observations were performed in Photon Counting mode \citep{2004SPIE.5165..217H}. The source counts were extracted in circular regions with radii of 47$''$ centered at the object, and the background counts were extracted from the source free circular regions with the same radii of 70$''$. We used the \texttt{XSELECT} to extract our spectra. We utilized the \texttt{XRTMKARF} tool to generate the ancillary response files for the 15 observations, and used the redistribution matrix files from the most recent \textit{Swift} calibration database (CALDB).   

There are six different band filters for the Ultraviolet and Optical Telescope (UVOT) \citep{2005SSRv..120...95R}: V, B, U, UVW1, UVM2 and UVW2. All of the UVOT observations have available data for at least one band. We chose circular regions with radii of 5$''$ to extract the source data and off-source circular regions with radii of 30$''$ to produce the background data. We used the tool \texttt{UVOTIMSUM} to sum up the sky images in a specific band filter and applied the \texttt{UVOTSOURCE} tool to perform aperture photometry in the summed up image for every band. Finally, we utilized the dust extinction function in \citet{1999PASP..111...63F} to perform the Galactic extinction correction for each UVOT filter with $\text{R}_\text{v}$ = 3.1 \citep{1999PASP..111...63F}, and the color excess of $\text{E}_\text{(B-V)}$ = 0.0066 \citep{2011ApJ...737..103S}.

\begin{figure*}
 \includegraphics[width=\textwidth]{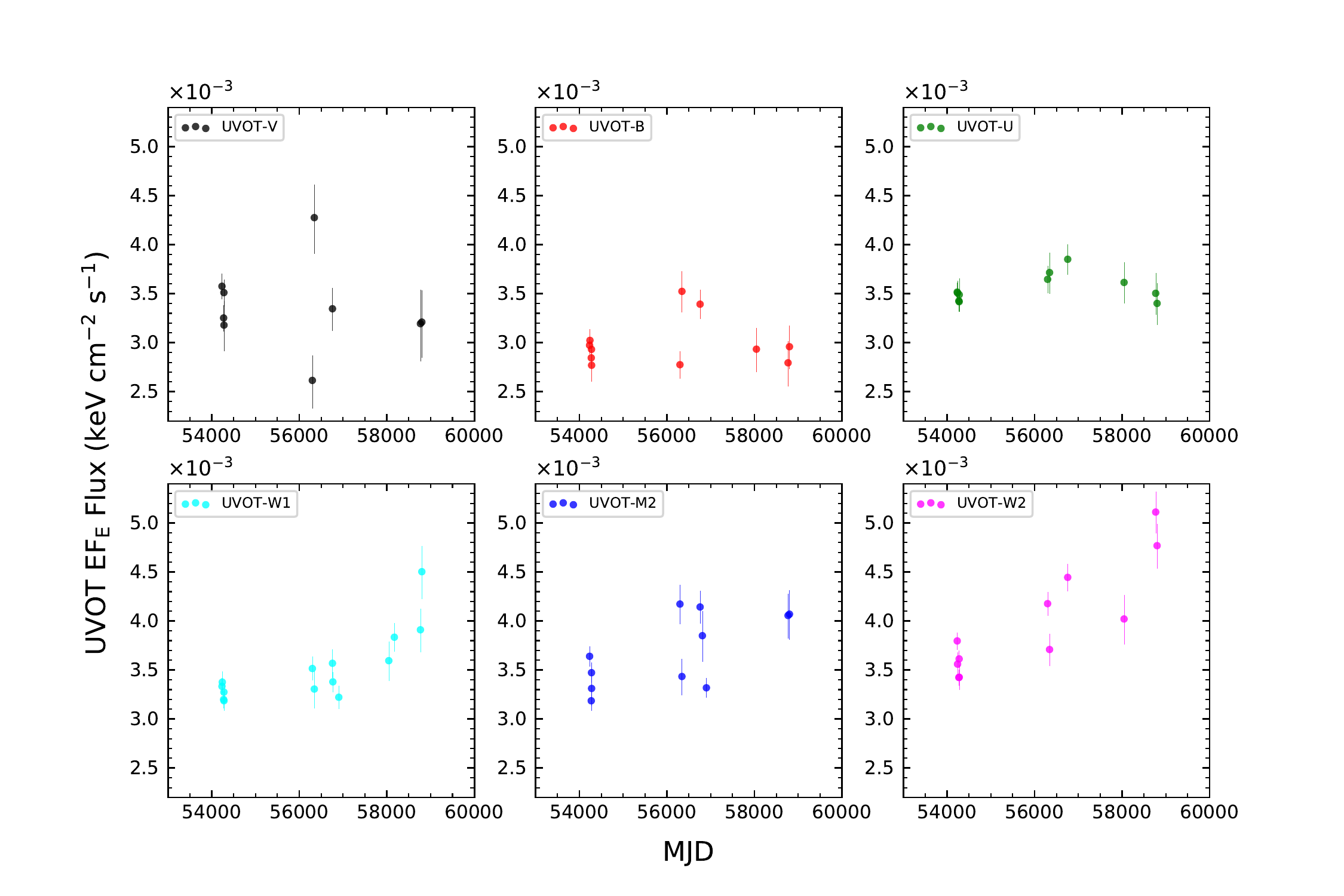}
 \vspace*{-7mm}
 \caption{The long-term \textit{Swift} UVOT light curves for six UV bands. The y-axes represent the monochromatic flux. The V band data are shown in black, B band in red, U band in green, UVW1 band in cyan, UVM2 band in blue and UVW2 band in magenta. The optical (V, B, U) monochromatic flux shows some random variations, while the UV (UVW1, UVM2, UVW2) flux shows a slight upward trend over a span of nearly 13 years.}
 \label{fig:UVOT}
\end{figure*}

\begin{figure*}
 \includegraphics[width=\textwidth]{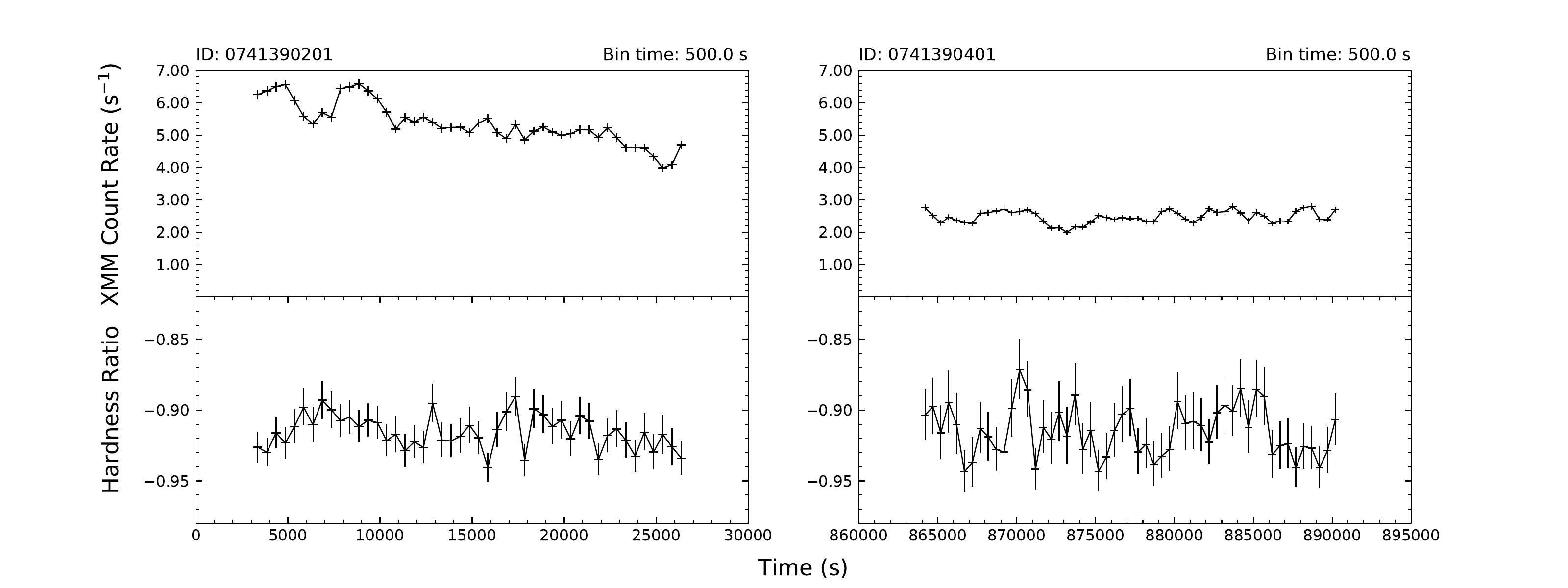}
  \vspace*{-5mm}
 \caption{Left panel: The short-term X-ray light curve and the corresponding hardness ratio of SBS 1353+564 for \textit{XMM-Newton} Obs. ID 0741390201. The bin time for the light curve is set to 500 s. The start time of the light curve is 56829 d 23:00:00 (MJD). The light curve shows an overall trend by decreasing from about 6 counts $\text{s}^{-1}$ down to 4 counts $\text{s}^{-1}$. Right panel: The light curve and the hardness ratio for Obs. ID 0741390401 with a bin time of 500 s. The start time is the same as that of Obs. ID 0741390201. The light curve is fairly consistent with a constant value of 2.4 counts $\text{s}^{-1}$ with some fluctuations. These two observations are ten days apart from each other. There is a clear drop in flux between the observations.}
 \label{fig:XMM}
\end{figure*}

\begin{figure}
 \centering
 \begin{subfigure}{.4\textwidth}
  \centering
  \includegraphics[width=\linewidth]{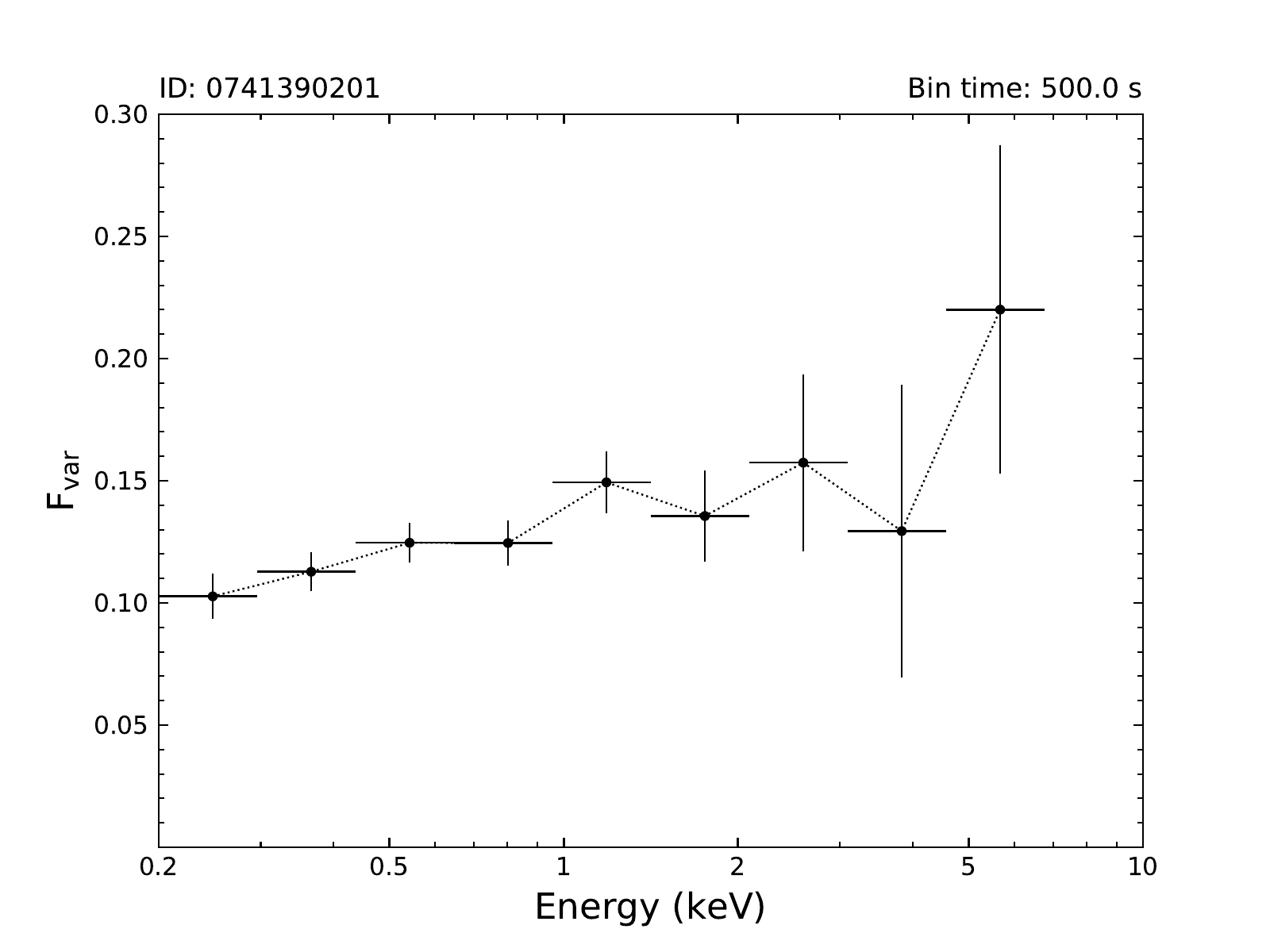}
 \end{subfigure}
 \begin{subfigure}{.4\textwidth}
  \centering
  \includegraphics[width=\linewidth]{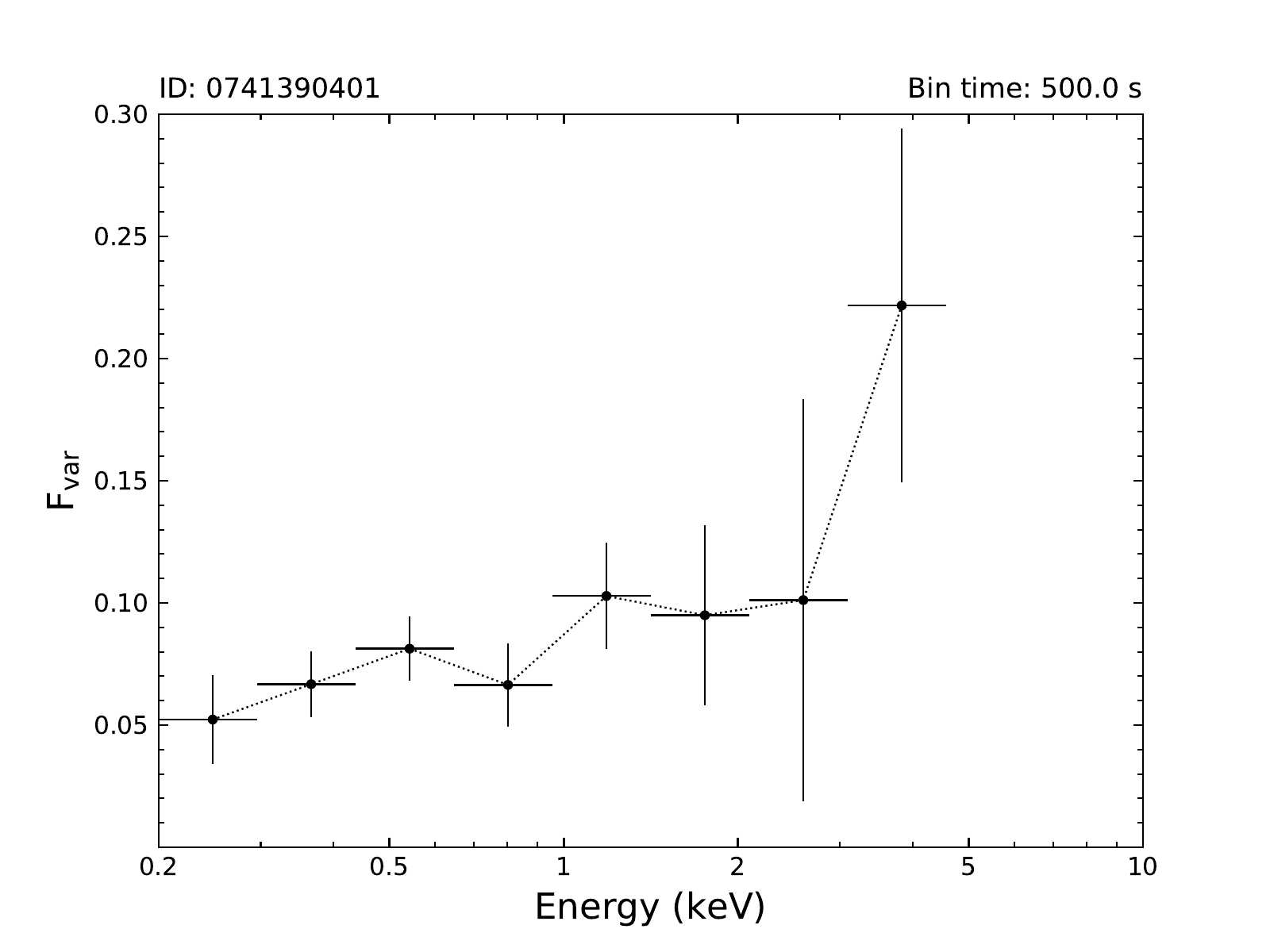}
 \end{subfigure}
 \vspace*{-1mm}                                                                          
 \caption{The fractional variability for \textit{XMM-Newton} observations. The bin time we used for extracting the light curve to calculate the $\textit{F}_\textit{var}$ is 500 s. The two fractional variabilities both show an uprise as the energy increases.}
 \label{fig:fvar}
\end{figure}

\section{Temporal Variability}

In this section we discuss the variability of SBS 1353+564 in the \textit{Swift} XRT/UVOT observations and the \textit{XMM-Newton} observations. We took into account the \textit{Swift} XRT data of a total of 15 observations \citep{2010ApJS..187...64G}. We extracted the XRT light curve using the 0.3 -- 10 keV count rate. Considering that the spectral data counts are low for the XRT instrument, we calculated the hardness ratio following the Bayesian method as described in \citet{2006ApJ...652..610P}. The soft and hard bands are referred to the 0.3 -- 2 keV and 2 -- 10 keV. We took into account the fractional difference hardness ratio, $\mathcal{HR}$, and chose the mean of the distribution as the estimation of $\mathcal{HR}$. The XRT light curve spanning nearly 13 years is plotted in Figure~\ref{fig:XRT}. The X-ray count rate ranges from 0.06 to 0.26 $\text{s}^{-1}$ presenting variabilities by a factor of $\sim$ 4, and the hardness ratio is relatively constant. We also considered the \textit{Swift} UVOT data for six different bands: V, B, U, UVW1, UVM2 and UVW2. All of the observations can output the UV/optical photometric data for at least one band. The UVOT light curve is shown in Figure~\ref{fig:UVOT}. The optical (V, B, U) flux exhibits no significant variability, while the UV (UVW1, UVM2, UVW2) flux shows a slight upward trend. From both \textit{Swift} XRT and UVOT observations over a 13-year timescale, there shows an obvious but random variation in SBS 1353+564, which is usually quite normal for NLS1s in the long-term timescale\citep{2010ApJS..187...64G}.

The short-term X-ray light curves and hardness ratios for \textit{XMM-Newton} Obs. ID 0741390201 and Obs. ID 0741390401 are shown in Figure~\ref{fig:XMM}. We make use of the 0.2 -- 10 keV count rate as our light curve, and the hardness ratio is calculated as $\mathcal{HR}$ = (H - S) / (H + S), where S and H are the source counts in the soft (0.2 -- 2 keV), and the hard (2 -- 10 keV) bands \citep{2006ApJ...652..610P}. The light curves have a bin size of 500s by the use of the \textit{lcurve} tool, and the x-axes are sharing the same start time: 56829 d 23:00:00 (MJD). The first observation light curve shows a general trend by decreasing from about 6 counts $\text{s}^{-1}$ down to 4 counts $\text{s}^{-1}$, while the second one is relatively consistent with a constant value of about 2.4 counts $\text{s}^{-1}$. The two observations are ten days apart from each other, which is relatively a short time in the long-term timescale. We can see there is a clear drop in flux, but the hardness ratio remains almost the same.

The fractional variability ($\textit{F}_\textit{var}$) is often used for analyzing the relative strength of the flux variability in each sliced energy band \citep{2002ApJ...568..610E, 2003MNRAS.345.1271V}. We sliced the 0.2 -- 10 keV energy band into ten logarithmically equally divided parts and processed the light curve with bin size of 500 s for each sliced band. We calculated the $\textit{F}_\textit{var}$ of the ten light curves. Figure~\ref{fig:fvar} shows the fractional variabilities of the two \textit{XMM-Newton} observations. We note that the $\textit{F}_\textit{var}$ in the 6.76 -- 10 keV band for Obs. ID 0741390201 and in the 4.57 -- 10 keV band for Obs. ID 0741390401 are absent because when calculating the $\textit{F}_\textit{var}$ amplitude in these bands, the total variance of the light curve ($\textit{S}^{2}$) is smaller than the mean error squared ($\langle{\sigma}^{2}_{err}\rangle$) \citep[see][Equation (1)]{2002ApJ...568..610E}, so by the definition we could not obtain the wanted amplitude. Although the two fractional variabilities exhibit an upward trend, they are both roughly comparable with two constants separately as the energy increases.

Overall, SBS 1353+564 only shows random variations in the long term of nearly 13 years. In the short term, the hardness ratios and the $\textit{F}_\textit{var}$ staying almost unchanged indicates that the X-ray spectra shape is practically invariable, while the light curves imply that the source experienced a flux drop between the two observations that are only 10 days apart.

\section{X-ray Spectral Analysis}
 
We used {\sevensize XSPEC} v.12.10.1f \citep{1996ASPC..101...17A} to analyze the \textit{XMM-Newton} spectra. We adopted the ${\chi}^{2}$ statistics to evaluate the goodness of fit. All errors of the fitted parameter values are derived using the 90\% confidence range (i.e., 2.706 ${\sigma}$ range). The Galactic absorption column density of $\textit{N}_\text{H} = 7.80\times10^{19}  \text{cm}^{-2}$ \citep{2016A&A...594A.116H} was fixed in all the fitting processes, and the \texttt{TBabs} model was used to account for it.

\begin{figure}
 \includegraphics[width=\columnwidth]{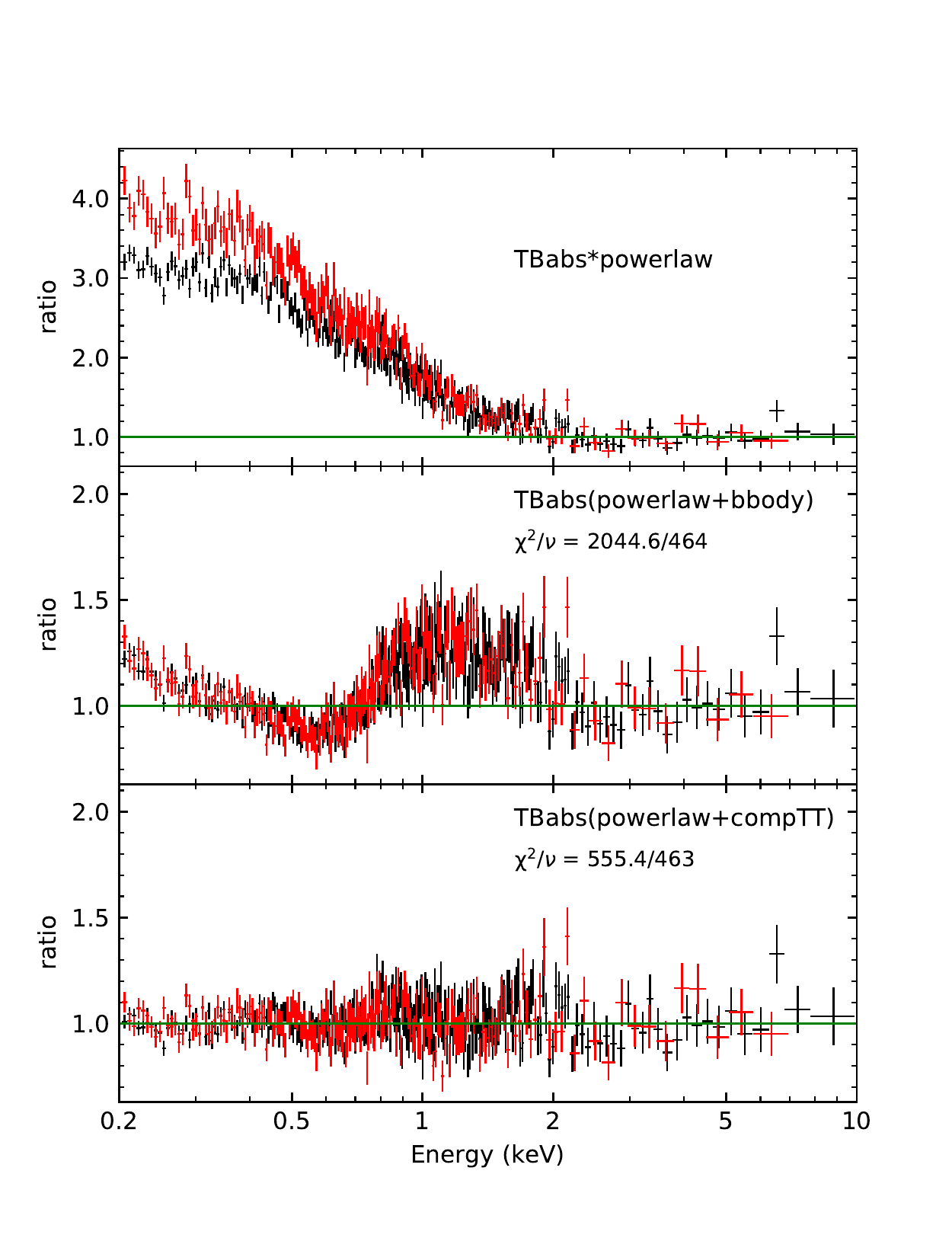}
 \vspace*{-10mm}
 \caption{The ratio plots of SBS 1353+564 for \textit{XMM-Newton} observations in 0.2 -- 10 keV band. Obs. ID 0741390201 and 0741390401 are plotted in black and red, respectively. A single power law modified by a Galactic absorption resulted in prominent residuals in the soft band, which can be better described by a Comptonization than a blackbody.}
 \label{fig:1}
\end{figure}

We started by fitting the 2 -- 10 keV energy band with a single power law modified by a Galactic absorption, and then extrapolated to the 0.2 -- 2 keV band. There is a prominent soft excess feature in both observations. We did not find any apparent Fe K emission line, but only noticed a minor positive residual between 6 -- 7 keV in the first observation. To account for the excess in the soft band, we then added a blackbody component to the power law in the 0.2 -- 10 keV. The power law parameters were frozen as the original fitted values, and the blackbody parameters were left free to vary. It turned out to be a very bad fit with ${\chi}^{2}/{\nu}$ = 2044.6/464. Because the blackbody model failed to fit the soft band, we tried to replace it with \texttt{compTT}. The \texttt{compTT} model \citep{1994ApJ...434..570T} describes a Comptonization which is often used to model the soft excess. This model resulted in a good fit with ${\chi}^{2}/{\nu}$ = 555.4/463. The data/model ratio plots of these three models are presented in Figure~\ref{fig:1}.

\begin{figure}
 \includegraphics[width=\columnwidth]{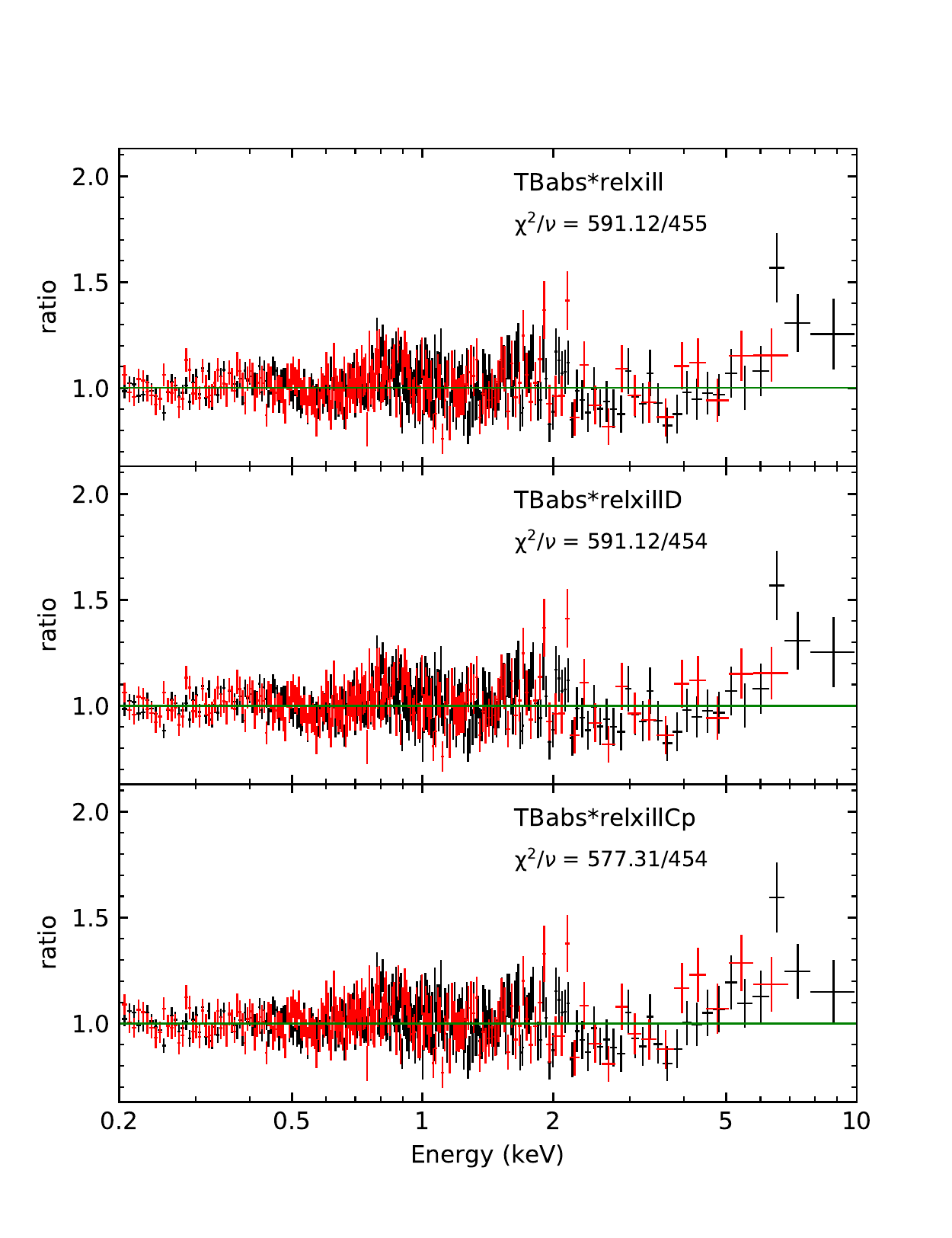}
 \vspace*{-10mm}
 \caption{The data/model ratio plots of SBS 1353+564 fitted with three different relativistic reflection models. The color scheme is the same as in Figure~\ref{fig:1}. These models generally fitted well but only left an excess feature in the high energy end above around 5 keV.}
 \label{fig:2}
\end{figure}

\begin{figure}
 \includegraphics[width=\columnwidth]{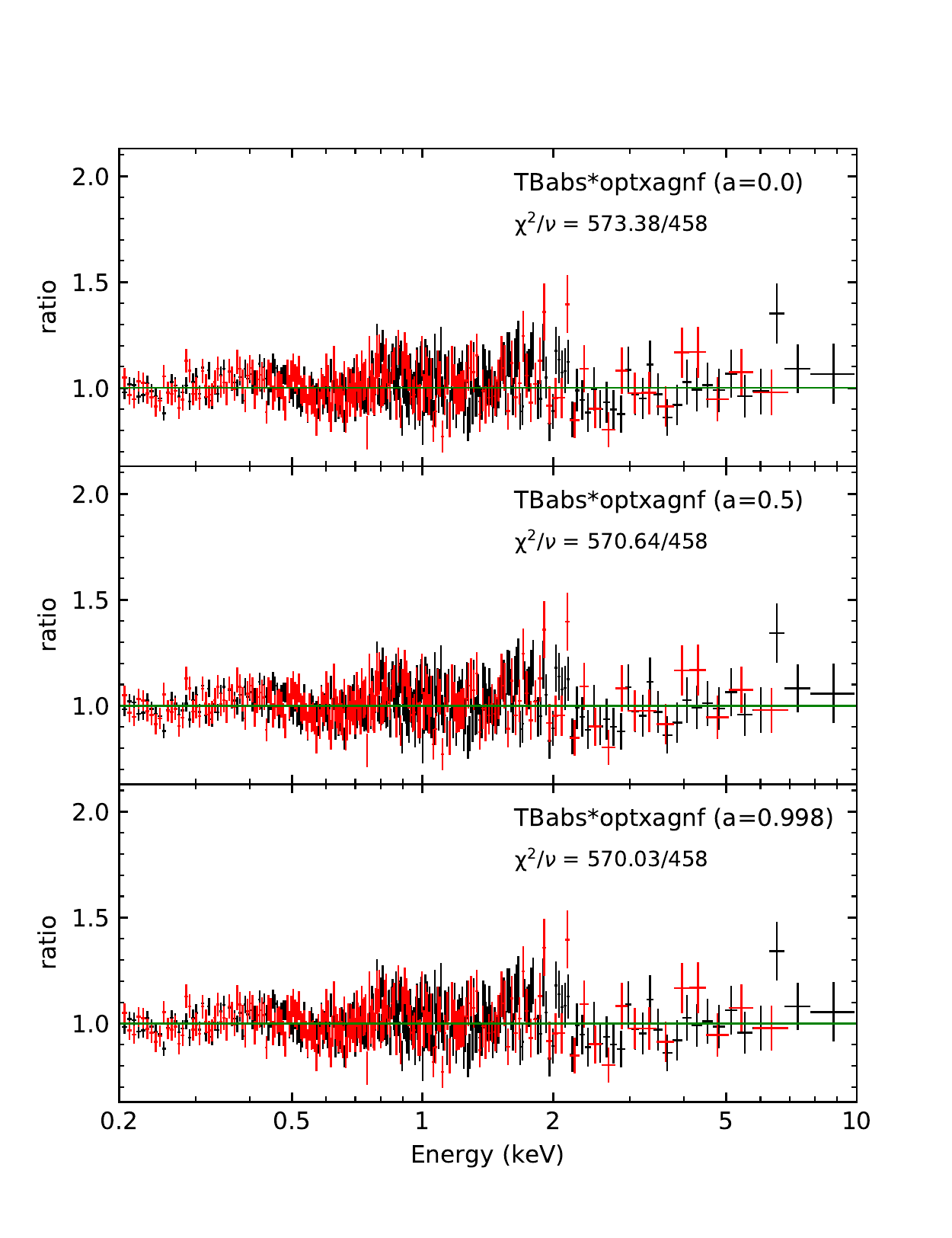}
 \vspace*{-10mm}
 \caption{The ratios of the best-fitted warm corona model \texttt{optxagnf} for three different spin parameters: a = 0.0, 0.5, 0.998. The color scheme stays the same as the previous figures. Except for the small peak between 6 and 7 keV in Obs. ID 0741390201, the fits are generally good for all the spins.}
 \label{fig:3}
\end{figure}

\subsection{Relativistically smeared reflection model}

Although a power law plus a Comptonization can describe the spectra pretty well, it cannot give a clear recognition of the physical scenario. To better understand the physical nature of the X-ray emission in SBS 1353+564, we use a set of relativistic reflection models. The model \texttt{relxill} \citep{2014ApJ...782...76G} combines reflection with relativistic smearing, which is an upgraded version of the neutral non-relativistic reflection model, \texttt{xillver} \citep{2013ApJ...768..146G}. This model contains a broken power law continuum as an incident spectrum, and a reflected spectrum which includes the soft X-ray excess. It assumes a geometry of an optically thin hot corona irradiating the surface of the ionized accretion disk with a hard X-ray continuum, and the hard X-ray photons are back-scattered from the surface, producing a relativistic reflection which often contributes to the soft excess, the broad Fe K${\alpha}$ emission line and sometimes the Compton hump. 

The relativistic reflection model has been applied to the X-ray spectral fitting of SBS 1353+564 before \citep{2020MNRAS.498.3888J}. \citet{2020MNRAS.498.3888J} applied the model \texttt{relconv} convolved with \texttt{reflionx} to the \textit{XMM-Newton} data of SBS 1353+564 accounting for the relativistic disk reflection component, and the model \texttt{nthcomp} accounting for the thermal Comptonization component. While we use three different sets of \texttt{relxill} model flavors: \texttt{relxill}, \texttt{relxillD} and \texttt{relxillCp} in our spectral fitting. In the \texttt{relxill} model, the parameter ${R}_{br}$ determines a break radius which separates two regimes with different emissivity profiles. The emissivity index between ${R}_{br}$ and ${R}_{out}$ (\textit{index}2) is fixed to be 3, considering that this Newtonian region is far from the inner extremely relativistic area, and the emissivity index between ${R}_{br}$ and ${R}_{in}$ (\textit{index}1) is freed to vary \citep{2018ApJ...855....3T, 2020MNRAS.497.4213G}. ${R}_{out}$ is fixed to be 500 gravitational radii and ${R}_{in}$ is set to be the ISCO (innermost stable circular orbit). We fix ${R}_{br}$ to be 6 gravitational radii for better constraint for other fitting parameters \footnote{The ISCO for a non-spinning black hole is 6 $\text{r}_{g}$ while it reduces to 1.27 $\text{r}_{g}$ as the spin parameter \textit{a} takes the maximal value of 0.998, and in extreme Kerr black hole case (\textit{a} = 1) the ISCO becomes 1 $\text{r}_{g}$ \citep{1974ApJ...191..507T, 2015PhRvD..91l4030J, 2020MNRAS.497.2352M}.}. The observed high energy cutoff, ${E}_{cut}$, of the primary spectrum is frozen to 300 keV, and the redshift to the source is set to \textit{z} = 0.1215 \citep{2017ApJS..229...39R}, all the rest parameters are variable. The model \texttt{relxillD} is the same as \texttt{relxill}, but allowing a higher density for the accretion disk varying from ${10}^{15}$ to ${10}^{19}$ $\text{cm}^{-3}$, and the high energy cutoff ${E}_{cut}$ was set to 300 keV under this model. The model \texttt{relxillCp} presumes a more physical scenario that the incident primary spectrum is an nthcomp Comptonization continuum \citep{1999MNRAS.309..561Z} derived from the corona where the thermal UV/optical photons produced in the disk are inverse Compton scattered to higher energies \citep{1980A&A....86..121S}.

Our results are basically consistent with those obtained by \citet{2020MNRAS.498.3888J}. We and \citet{2020MNRAS.498.3888J} both derive good fits for the X-ray spectra of SBS 1353+564. The data/model ratios of our results are plotted in Figure~\ref{fig:2}, and the results are shown in Table~\ref{tab:1}, where we can find that the best-fitting parameters between the two observations are similar except for the normalization. Our results show that the black hole has a rapid spin with the spin parameter a > 0.9, which agrees with the analysis by \citet{2020MNRAS.498.3888J}. The disc inclination given by \citet{2020MNRAS.498.3888J} is a high value with \textit{i} $\sim$ 60 deg, which is similar to the inclination obtained under our \texttt{relxill} and \texttt{relxillD} models with \textit{i} = $56\pm8$ deg. Besides, both of us find that the source SBS 1353+564 shows a very high reflection fraction, which means the reflection component contributing to the soft excess is quite significant in the X-ray band. However, the reflection fraction derived from \citet{2020MNRAS.498.3888J} is $\sim$ 1.0, while we obtain a larger value of $\sim$ 2.0 for $\mathcal{R}_\text{refl}$. The iron abundance is moderate with $2.6\pm1.1$ $\text{A}_{\odot}$ obtained under our \texttt{relxill} and \texttt{relxillD} models, and $1.4\pm0.4$ $\text{A}_{\odot}$ under our \texttt{relxillCp} model, and \citet{2020MNRAS.498.3888J} gave $2.0\pm0.2$ $\text{A}_{\odot}$ basically agreeing with our results. As for the ionization of the accretion disk, we and \citet{2020MNRAS.498.3888J} both give the ionization around log$({\xi})$ $\gtrsim$ 3, which can be considered as a heavily ionized state. Our \texttt{relxillD} model allowing the disk density varying from ${10}^{15}$ to ${10}^{19}$ $\text{cm}^{-3}$ does not improve the fitting compared with the \texttt{relxill} model with a fixed density of ${10}^{15}$ $\text{cm}^{-3}$. Both the \texttt{relxill} and \texttt{relxillD} models have almost the same parameter values and ${\chi}^{2}$ as shown in Table~\ref{tab:1}, and the density of the accretion disk under the \texttt{relxillD} model is pegged at ${10}^{15}$ $\text{cm}^{-3}$, which is consistent with the result obtained by \citet{2020MNRAS.498.3888J} that the disk density has an upper limit (i.e., log $N$/$\text{cm}^{-3}$ < 16.2), indicating that the disk density of the source is quite low. The power law index of the incident spectrum by \citet{2020MNRAS.498.3888J} is $\sim$ 2.490, agreeing with the values of the index (i.e., ${\Gamma}$ $\sim$ 2.5) under the \texttt{relxill} and \texttt{relxillD} models, indicating that the continuum is quite soft.

Although our results mainly agree with the results by \citet{2020MNRAS.498.3888J}, the data/model ratio plots are slightly different. \citet{2020MNRAS.498.3888J} found no obvious structural residuals in their plots using the spectra between 0.5 and 10 keV for reflection model fitting. Our models fitting the spectra between 0.2 and 10 keV generally fitted well, however, there are some positive residuals above around 5 keV as can be seen in Figure~\ref{fig:2}. Moreover, there is a small peak between 6 and 7 keV in Obs. ID 0741390201 which maybe can be attributed to an iron emission line. For the two observations, there is practically no variability in spectral shape, which only differ in flux. The variability in flux indicates that the source SBS 1353+564 experienced a flux decrease by a factor of $\sim$ 2 in merely ten days.

\begin{table*}
 \caption{The best-fitting model parameters for the relativistic reflection models of SBS 1353+564 for \textit{XMM-Newton} observations (0.2 -- 10 keV). The superscript `${f}$' denotes the fixed parameters. The superscript `${t}$' denotes the tied parameters between observations. The superscript `${p}$' denotes the parameters pegged to the upper or lower limit. }
 \label{tab:1}
 \begin{tabular}{ c c c c c c }
   \hline
   Date &   &   &   & 2014-06-21 & 2014-07-01\\
   \hline
   Obs. ID &   &   &   & 0741390201 & 0741390401\\
   \hline
   Model & Model Component & Model Parameter & Unit &   &   \\
   \hline
   Relativistic reflection & \texttt{TBabs} & $\textit{N}_{H}$ & $\text{10}^{19}$ $\text{cm}^{-2}$ & ${7.8}^{f}$ & ${7.8}^{f}$\\
   (Relxill) & \texttt{relxill} & \textit{index}1 &   & $8.8\pm1.9$ & ${10}^{p}$\\
      &   & \textit{index}2 &   & ${3}^{f}$ & ${3}^{f}$\\
      &   & \textit{a} &   & ${0.99\pm0.01}^{t}$ & ${0.99\pm0.01}^{t}$\\
      &   & \textit{i} & deg & ${56\pm8}^{t}$ & ${56\pm8}^{t}$\\
      &   & ${\Gamma}$ &   & ${2.55\pm0.03}$ & ${2.54\pm0.03}$\\
      &   & log$({\xi})$ & log(erg cm $\text{s}^{-1}$) & ${3.2\pm0.3}$ & ${3.2\pm0.2}$\\
      &   & $\textit{A}_\textit{Fe}$ & $\text{A}_{\odot}$ & ${2.6\pm1.1}^{t}$ & ${2.6\pm1.1}^{t}$\\
      &   & $\textit{E}_\textit{cut}$ & keV & ${300}^{f}$ & ${300}^{f}$\\
      &   & $\textit{Norm}$ & ${10}^{-5}$ & ${1.0\pm0.5}$ & ${0.4\pm0.2}$\\
      & Fit Quality & ${\chi}^{2}/{\nu}$ &   & \multicolumn{2}{c}{591.12/455}\\
   \hline
   Relativistic reflection & \texttt{TBabs} & $\textit{N}_{H}$ & $\text{10}^{19}$ $\text{cm}^{-2}$ & ${7.8}^{f}$ & ${7.8}^{f}$\\
   (RelxillD) & \texttt{relxillD} & \textit{index}1 &   & $8.8\pm1.9$ & ${10}^{p}$\\
      &   & \textit{index}2 &   & ${3}^{f}$ & ${3}^{f}$\\
      &   & \textit{a} &   & ${0.99\pm0.01}^{t}$ & ${0.99\pm0.01}^{t}$\\
      &   & \textit{i} & deg & ${56\pm8}^{t}$ & ${56\pm8}^{t}$\\
      &   & ${\Gamma}$ &   & ${2.55\pm0.03}$ & ${2.54\pm0.04}$\\
      &   & log$({\xi})$ & log(erg cm $\text{s}^{-1}$) & ${3.2\pm0.3}$ & ${3.2\pm0.2}$\\
      &   & $\textit{A}_\textit{Fe}$ & $\text{A}_{\odot}$ & ${2.6\pm1.1}^{t}$ & ${2.6\pm1.1}^{t}$\\
      &   & log $N$ & log($\text{cm}^{-3}$) & ${15}^{t,p}$ & ${15}^{t,p}$\\
      &   & $\textit{Norm}$ & ${10}^{-5}$ & ${1.0\pm0.5}$ & ${0.4\pm0.3}$\\
      & Fit Quality & ${\chi}^{2}/{\nu}$ &   & \multicolumn{2}{c}{591.12/454}\\
   \hline
   Relativistic reflection & \texttt{TBabs} & $\textit{N}_{H}$ & $\text{10}^{19}$ $\text{cm}^{-2}$ & ${7.8}^{f}$ & ${7.8}^{f}$\\
   (RelxillCp) & \texttt{relxillCp} & \textit{index}1 &   & $7.9\pm0.5$ & $8.5\pm0.6$\\ 
      &   & \textit{index}2 & & ${3}^{f}$ & ${3}^{f}$\\
      &   & \textit{a} &   & ${0.91\pm0.06}^{t}$ & ${0.91\pm0.06}^{t}$\\
      &   & \textit{i} & deg & ${29\pm7}^{t}$ & ${29\pm7}^{t}$\\
      &   & ${\Gamma}$ &   & ${2.19\pm0.06}$ & ${2.21\pm0.08}$\\
      &   & log$({\xi})$ & log(erg cm $\text{s}^{-1}$) & ${3.0\pm0.1}$ & ${3.0\pm0.2}$\\
      &   & $\textit{A}_\textit{Fe}$ & $\text{A}_{\odot}$ & ${1.4\pm0.4}^{t}$ & ${1.4\pm0.4}^{t}$\\
      &   & $\textit{k}{T}_{e}$ & $\text{keV}$ & ${4.8\pm0.7}^{t}$ & ${4.8\pm0.7}^{t}$\\
      &   & $\textit{Norm}$ & ${10}^{-5}$ & ${0.2\pm0.2}$ & ${0.1\pm0.1}$\\
      & Fit Quality & ${\chi}^{2}/{\nu}$ &   & \multicolumn{2}{c}{577.31/454}\\
   \hline   
 \end{tabular} 
\end{table*}

\subsection{Warm corona model}
\label{sec:warmcorona}

As we discussed in Section 4.1, the relativistic reflection model can generally fit the spectra except for the high energy end above around 5 keV. Hence, we carry on another physical model to explain the nature of X-ray in SBS 1353+564. The second model we used is the warm corona model \texttt{optxagnf} \citep{2012MNRAS.420.1848D}. This model assumes three components powered by gravitational energy released in accretion. The first component is the UV/optical emission as a blackbody from the outer region of the color temperature corrected disk down to the coronal radius, $\textit{R}_{cor}$. The second is an optically thick, low temperature (warm) corona where the UV/optical photons are inverse Compton scattered to the soft X-ray band producing the soft X-ray excess. The third is an optically thin, high temperature (hot) corona which produces the hard X-ray Comptonized power law continuum. During the fitting procedure, the model normalisations were frozen to unity because the flux is set by the parameters of the black hole mass, the Eddington ratio and the black hole spin. As a parameter in \texttt{optxagnf}, the central black hole mass is calculated using the virial mass estimate expression and calibration for $\lambda = 5100$ $\si{\angstrom}$ and H${\beta}$ \citep[see][Equation (5)]{2006ApJ...641..689V}. We obtain the emission line properties of SBS 1353+564 from \citet[Table 1]{2017ApJS..229...39R}. ${L}_\text{\sevensize5100\si{\angstrom}}$ is estimated to be 1.10 $\times$ ${10}^{44}$ erg $\text{s}^{-1}$, and $\text{FWHM}_{H\beta}$ is measured to be 2126 km $\text{s}^{-1}$, which is slightly larger than 2000 km $\text{s}^{-1}$, defined as the upper limit for NLS1s. The black hole mass determined by the equation described as \citet[Equation (5)]{2006ApJ...641..689V} using ${L}_\text{\sevensize5100\si{\angstrom}}$ and $\text{FWHM}_{H\beta}$ is calculated to be $\sim$ 3.8 $\times$ ${10}^{7}$ $\text{M}_{\odot}$, which is input in \texttt{optxagnf} and frozen throughout the fitting. Further discussion about the determination of the black hole mass will be given in the section~\ref{sec:blackholemass}. The outer radius of the disk is frozen to ${10}^{5}$ $\text{r}_{g}$. The coronal radius and the black hole spin parameter could not be well constrained when set as free parameters. Hence, we fixed the coronal radius, $\textit{R}_{cor}$, to 20 $\text{r}_{g}$ in the fitting process. For the spin parameter, we tried three values of spin: 0.0, 0.5 and 0.998, which correspond to non-spin, intermediate spin and maximal spin scenarios, respectively. The best-fit with these three different spin values are ${\chi}^{2}/{\nu}$ = 573.38/458 for a $\sim$ 0.0, ${\chi}^{2}/{\nu}$ = 570.64/458 for a $\sim$ 0.5 and ${\chi}^{2}/{\nu}$ = 570.03/458 for a $\sim$ 0.998, which can all be considered as good fits to the X-ray band spectra. These fit statistics do not deviate too much from each other, but the best-fitting parameters in the three cases are not very alike, especially the Eddington ratio. With the spin parameter increasing, the fitted Eddington ratio becomes smaller for each observation. Between the two observations, the early one (Obs. ID 0741390201) has a larger Eddington ratio than the later one (Obs. ID 0741390401). The fitting results are listed in Table~\ref{tab:2} and the best-fit model ratios are plotted in Figure~\ref{fig:3}.

\begin{table*}
 \caption{The best-fitting parameters for the warm corona model at three different spin values in 0.2 -- 10 keV band of SBS 1353+564. The superscript `${f}$' denotes the frozen parameters during fitting. These three sets of parameters vary from each other, but the fit statistics are roughly the same.}
 \label{tab:2}
 \begin{tabular}{ c c c c c c }
   \hline
   Date &   &   &   & 2014-06-21 & 2014-07-01\\
   \hline
   Obs. ID &   &   &   & 0741390201 & 0741390401\\
   \hline
   Model & Model Component & Model Parameter & Unit &   &   \\
   \hline
   \vspace*{+1mm}
   Warm corona & \texttt{TBabs} & $\textit{N}_{H}$ & $\text{10}^{19}$ $\text{cm}^{-2}$ & ${7.8}^{f}$ & ${7.8}^{f}$\\
   \vspace*{+1mm}
   (a = 0.0) & \texttt{optxagnf} & log$\textit({L}_{bol}/{L}_{Edd})$ &   & ${-0.284}^{+0.015}_{-0.015}$ & ${-0.532}^{+0.024}_{-0.025}$\\
   \vspace*{+1mm}
      &   & $\textit{R}_{cor}$ & $\text{r}_{g}$ & ${20}^{f}$ & ${20}^{f}$\\
   \vspace*{+1mm}
      &   & $\textit{k}{T}_{e}$ & keV & ${0.22}^{+0.02}_{-0.02}$ & ${0.21}^{+0.03}_{-0.03}$\\
   \vspace*{+1mm} 
      &   & ${\tau}$ &   & ${16.3}^{+1.6}_{-1.4}$ & ${16.4}^{+2.1}_{-1.8}$\\
   \vspace*{+1mm}   
      &   & ${\Gamma}$ &   & ${2.31}^{+0.08}_{-0.08}$ & ${2.26}^{+0.11}_{-0.12}$\\
   \vspace*{+1mm}   
      &   & $\textit{f}_{pl}$ &   & ${0.39}^{+0.02}_{-0.03}$ & ${0.32}^{+0.08}_{-0.06}$\\
      & Fit Quality & ${\chi}^{2}/{\nu}$ &   & \multicolumn{2}{c}{573.38/458}\\
   \hline
   \vspace*{+1mm}
   Warm corona & \texttt{TBabs} & $\textit{N}_{H}$ & $\text{10}^{19}$ $\text{cm}^{-2}$ & ${7.8}^{f}$ & ${7.8}^{f}$\\
   \vspace*{+1mm}
   (a = 0.5) & \texttt{optxagnf} & log$\textit({L}_{bol}/{L}_{Edd})$ &   & ${-0.403}^{+0.015}_{-0.016}$ & ${-0.651}^{+0.028}_{-0.029}$\\
   \vspace*{+1mm}
      &   & $\textit{R}_{cor}$ & $\text{r}_{g}$ & ${20}^{f}$ & ${20}^{f}$\\
   \vspace*{+1mm}   
      &   & $\textit{k}{T}_{e}$ & keV & ${0.22}^{+0.03}_{-0.02}$ & ${0.21}^{+0.03}_{-0.03}$\\
   \vspace*{+1mm}   
      &   & ${\tau}$ &   & ${15.9}^{+1.6}_{-1.4}$ & ${16.3}^{+2.1}_{-1.8}$\\
   \vspace*{+1mm}   
      &   & ${\Gamma}$ &   & ${2.30}^{+0.08}_{-0.08}$ & ${2.26}^{+0.11}_{-0.12}$\\
   \vspace*{+1mm}   
      &   & $\textit{f}_{pl}$ &   & ${0.36}^{+0.07}_{-0.05}$ & ${0.31}^{+0.08}_{-0.06}$\\
      & Fit Quality & ${\chi}^{2}/{\nu}$ &   & \multicolumn{2}{c}{570.64/458}\\
   \hline
   \vspace*{+1mm}
   Warm corona & \texttt{TBabs} & $\textit{N}_{H}$ & $\text{10}^{19}$ $\text{cm}^{-2}$ & ${7.8}^{f}$ & ${7.8}^{f}$\\
   \vspace*{+1mm}
   (a = 0.998) & \texttt{optxagnf} & log$\textit({L}_{bol}/{L}_{Edd})$ &   & ${-0.547}^{+0.025}_{-0.026}$ & ${-0.795}^{+0.042}_{-0.026}$\\
   \vspace*{+1mm}
      &   & $\textit{R}_{cor}$ & $\text{r}_{g}$ & ${20}^{f}$ & ${20}^{f}$\\
   \vspace*{+1mm}   
      &   & $\textit{k}{T}_{e}$ & keV & ${0.23}^{+0.03}_{-0.02}$ & ${0.21}^{+0.03}_{-0.03}$\\
   \vspace*{+1mm}   
      &   & ${\tau}$ &   & ${15.8}^{+1.6}_{-1.4}$ & ${16.3}^{+2.1}_{-1.8}$\\
   \vspace*{+1mm}   
      &   & ${\Gamma}$ &   & ${2.30}^{+0.08}_{-0.09}$ & ${2.26}^{+0.11}_{-0.12}$\\
   \vspace*{+1mm}   
      &   & $\textit{f}_{pl}$ &   & ${0.31}^{+0.08}_{-0.06}$ & ${0.26}^{+0.10}_{-0.07}$\\
      & Fit Quality & ${\chi}^{2}/{\nu}$ &   & \multicolumn{2}{c}{570.03/458}\\
   \hline
 \end{tabular}
\end{table*}

\begin{figure}
 \includegraphics[width=\columnwidth]{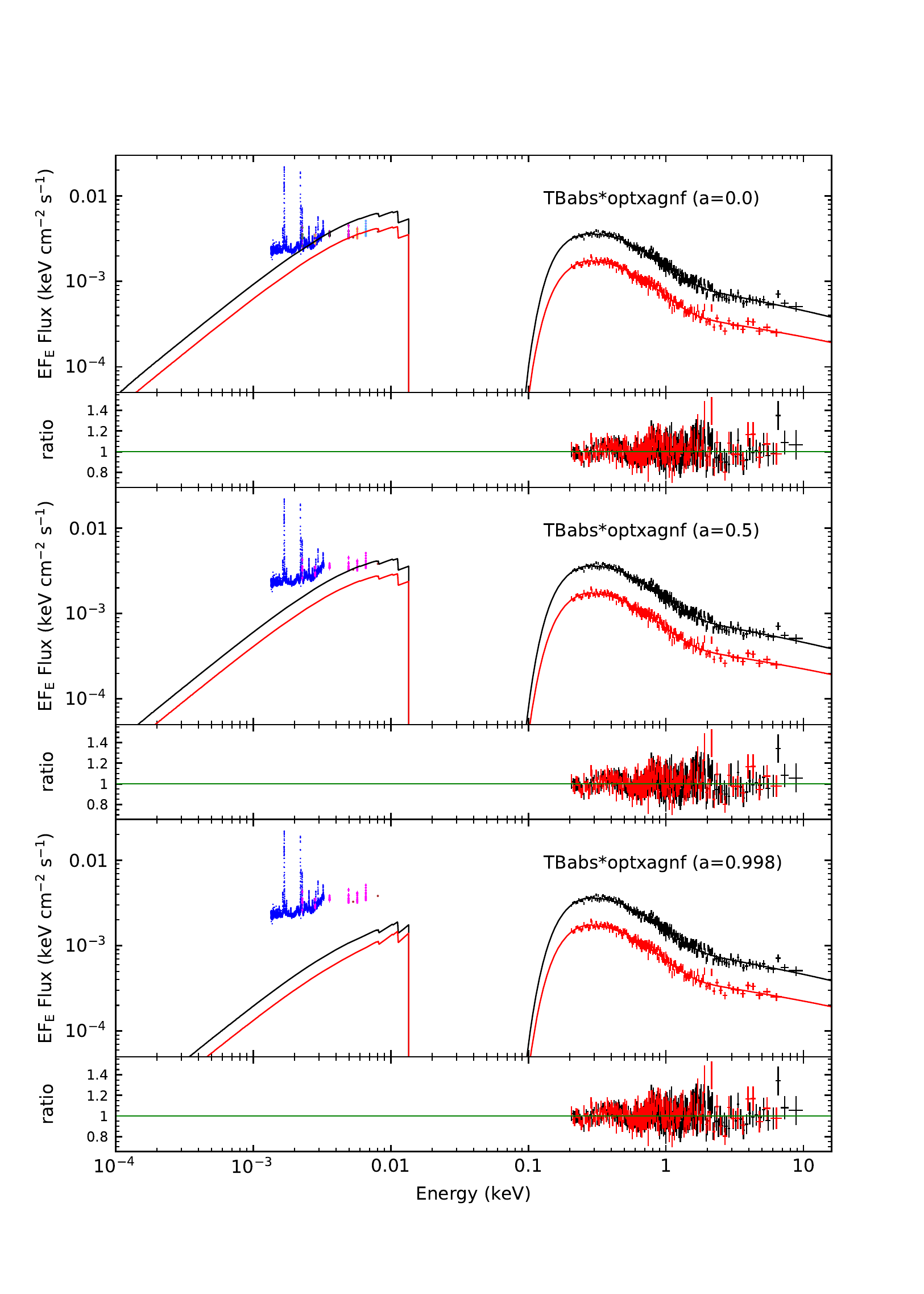}
 \vspace*{-13mm}
 \caption{The \textit{XMM-Newton} X-ray spectra of SBS 1353+564 with UV/optical spectral data from SDSS, \textit{Swift} UVOT and GALEX. The X-ray spectra are superimposed by the best-fit warm corona model \texttt{optxagnf} and the model is extrapolated to the lower energy band. The data/model ratio of each spectrum is shown below the spectrum. Obs. ID 0741390201 and 0741390401 are plotted in black and red, SDSS, \textit{Swift} UVOT and GALEX spectral data are shown in blue, magenta and brown, respectively. }
 \label{fig:4}
\end{figure}

In order to find out which set of the parameter values is the proper one for our source SBS 1353+564, we further considered the UV/optical data from other observations. Considering that the model \texttt{optxagnf} has a thermal disk component emitting UV/optical photons, we attempted to add some UV/optical data to our X-ray spectra. We took into account the SDSS optical spectrum of SBS 1353+564, along with the processed \textit{Swift} UVOT data in the V, B, U, UVW1, UVM2 and UVW2 bands for 16 observations over four years, and two GALEX data in the FUV and NUV bands \citep{2012AAS...21934001S}. We added these UV/optical data to our \textit{XMM-Newton} X-ray spectra and extrapolate our best-fit models to the lower energy band to compare the data and models. The extended spectra and models are displayed in Figure~\ref{fig:4}, where we can see that the extrapolated models exhibit a gap in the EUV band (10.25 eV -- 124 eV), which is attributed to Galactic absorption \citep{2000ApJ...542..914W}. The UV/optical spectral data from SDSS, \textit{Swift} UVOT and GALEX are plotted in blue, magenta and brown, respectively. We can see that in the SDSS spectra the H${\alpha}$ and the [O {\sevensize III}] 5007 $\si{\angstrom}$ emission lines are quite obvious. As for the \textit{Swift} UVOT, there is a total of 16 observations for each band with only some random variations, and the V and B bands are shown overlapped with the SDSS spectra. Since the X-ray spectra are fitted almost equally well for all the spin values taken and the long-term UV/optical variability of this source is not statistically significant, we can determine the best-fit by the use of lower energy spectra. In these three different spin scenarios, Figure~\ref{fig:4} shows that the extrapolated model part (i.e., the disk component) can describe the UV/optical spectral data better at spin of $\sim$ 0.0 than $\sim$ 0.5 and $\sim$ 0.998. Therefore, we think the set of parameters achieved under the non-spin scenario is the best for our warm corona model. In the case of a $\sim$ 0.0, the temperature and the optical depth of this warm corona are $\textit{k}{T}_{e} \sim 220 $ eV and ${\tau} \sim 16$. The coronal radius, $\textit{R}_{cor}$, is set to 20 $\text{r}_{g}$, below which the gravitational energy is released in the hot and warm coronas. The fraction of the energy below $\textit{R}_{cor}$ which is emitted in the hot corona is $\textit{f}_{pl} \sim$ 0.3 -- 0.4, and the fraction emitted in the warm corona is 1 - $\textit{f}_{pl}$. As for the Eddington ratio, the fitted values of the two observations are log$\textit({L}_{bol}/{L}_{Edd}) = {-0.284}^{+0.015}_{-0.015}$ and ${-0.532}^{+0.024}_{-0.025}$. So far, from all the models applied, we found that the ${\chi}^{2}$ of the two models are quite close, however, the relativistically smeared reflection model is unable to fit the data above 5 keV well, while the warm corona model with a non-spin case can account for the soft X-ray excess in SBS 1353+564.

\section{Discussion and Conclusion}

SBS 1353+564 is classified as a Narrow Line Seyfert 1 galaxy and exhibits unique features in X-ray and UV/optical bands which often emerge in the cases of NLS1s. We performed for the first time the timing and spectral analyses for SBS 1353+564 using \textit{XMM-Newton} and \textit{Swift} long-term and multi-epoch X-ray data to investigate the nature of its soft X-ray excess. For the two \textit{XMM-Newton} observations, the EPIC-pn data were only considered because of their better signal to noise ratio than EPIC-MOS, and we retained the 0.2 -- 10 keV energy band for our model fitting. SBS 1353+564 was also observed by \textit{Swift} for 15 times, providing available data for XRT and UVOT. Using the EPIC-pn spectra in 0.2 -- 10 keV band, we simultaneously fitted the spectra with two sets of physical models: relativistic reflection and warm corona models. It resulted that the two models can generally yield good fits, but the warm corona model is more preferred in explaining the overall spectra.

\subsection{Black hole properties}
\label{sec:blackholemass}

In the section~\ref{sec:warmcorona}, we estimate the black hole mass using the virial mass expression \citep[see][Equation (5)]{2006ApJ...641..689V} with ${L}_\text{\sevensize5100\si{\angstrom}}$ and $\text{FWHM}_{H\beta}$ from \citet[Table 1]{2017ApJS..229...39R}, and our estimation is $\sim$ 3.8 $\times$ ${10}^{7}$ $\text{M}_{\odot}$. However, \citet{2010ApJS..187...64G} uses the method described in \citet{2000ApJ...533..631K} with ${L}_\text{\sevensize5100\si{\angstrom}}$ and $\text{FWHM}_{H\beta}$ from \citet{2004AJ....127..156G} and derives a mass of $\sim$ 5.4 $\times$ ${10}^{6}$ $\text{M}_{\odot}$. The black hole mass estimate methods in \citet{2006ApJ...641..689V} and \citet{2000ApJ...533..631K} can derive similar results with the same ${L}_\text{\sevensize5100\si{\angstrom}}$ and $\text{FWHM}_{H\beta}$ values, and here we use the recent method in \citet{2006ApJ...641..689V}, which is more improved in determining black hole mass. As for ${L}_\text{\sevensize5100\si{\angstrom}}$, \citet{2017ApJS..229...39R} and \citet{2004AJ....127..156G} give similar values around ${10}^{44}$ erg $\text{s}^{-1}$. However, $\text{FWHM}_{H\beta}$ from these two papers are quite different. \citet{2004AJ....127..156G} gives 1100 km $\text{s}^{-1}$, while \citet{2017ApJS..229...39R} gives 2126 km $\text{s}^{-1}$. To determine the more appropriate $\text{FWHM}_{H\beta}$ for our mass estimation, we roughly measure $\text{FWHM}_{H\beta}$ using the publicly available code PyQSOFit for spectral decomposition \citep{2018ascl.soft09008G} and the SDSS optical spectrum of SBS 1353+564. We obtain a value of 1.35 $\times$ ${10}^{44}$ erg $\text{s}^{-1}$ for ${L}_\text{\sevensize5100\si{\angstrom}}$. For $\text{FWHM}_{H\beta}$, we first fit the H${\beta}$ emission line with a single Gaussian component, and FWHM of the fitted Gaussian line is $\sim$ 1724 km $\text{s}^{-1}$. Next, we use two Gaussian lines to account for the H${\beta}$ broad and narrow components, and we derive a value of 2823 km $\text{s}^{-1}$ for FWHM of the H${\beta}$ broad component, which is a bit larger than 2126 km $\text{s}^{-1}$ given by \citet{2017ApJS..229...39R} using the SDSS spectrum as well. Our rough measurement of $\text{FWHM}_{H\beta}$ only helps us to determine the more appropriate value, while \citet{2017ApJS..229...39R} carried out a systematic and detailed spectral analysis, which gives a more accurate result. Hence, we use 2126 km $\text{s}^{-1}$ as $\text{FWHM}_{H\beta}$ for black hole mass estimation.

In order to find out what kind of effect the black hole mass has on the fitting process. We also calculate the mass using 2823 km $\text{s}^{-1}$ as $\text{FWHM}_{H\beta}$ and 1.35 $\times$ ${10}^{44}$ erg $\text{s}^{-1}$ as ${L}_\text{\sevensize5100\si{\angstrom}}$ with the same mass estimate method. We derive a mass estimation of $\sim$ 7.5 $\times$ ${10}^{7}$ $\text{M}_{\odot}$, which is used again in \texttt{optxagnf} for another spectral fitting. We also consider three different black hole spin scenarios, and the best-fit are ${\chi}^{2}/{\nu}$ = 575.42/456 for a $\sim$ 0.0, ${\chi}^{2}/{\nu}$ = 572.67/456 for a $\sim$ 0.5 and ${\chi}^{2}/{\nu}$ = 571.78/456 for a $\sim$ 0.998. These fits are quite alike so we once again extrapolate the best-fit models to the lower energy band to compare with those UV/optical data mentioned earlier. This time we find the extrapolated model can describe the UV/optical data better at spin of $\sim$ 0.5 than $\sim$ 0.0 and $\sim$ 0.998. Hence, we find the fitted spin parameter relies on the black hole mass that we input. The input black hole mass only varies by a factor of $\sim$ 2, while the fitted black hole spin changes from the non-spin to the intermediate spin. For the two cases with different masses, the extrapolated model parts only deviate a little from the UV/optical data at spin of $\sim$ 0.0 and $\sim$ 0.5, but they both deviate a lot from the UV/optical data at spin of $\sim$ 0.998, therefore, we guess that the central black hole in SBS 1353+564 is not an extreme rotating black hole. We cannot decide the black hole spin for sure but only guess that the black hole lies between the non-spin and the intermediate spin scenarios. However, in our case that the mass is taken as $\sim$ 3.8 $\times$ ${10}^{7}$ $\text{M}_{\odot}$, we still take a $\sim$ 0.0 as our spin parameter for further analysis.

\begin{table*}
 \caption{The UV and the X-ray properties of the two \textit{XMM-Newton} observations modelled by \texttt{optxagnf} (a = 0.0).}
 \label{tab:3}
 \begin{tabular}{ c c c c c c }
  \hline
  Obs. ID & $\text{log}({L}_\text{2500\si{\angstrom}})$ & $\text{log}({L}_\text{2 keV})$ & ${\alpha}_\text{\tiny OX}$ & $\text{log}({L}_\text{bol})$ & ${\lambda}_\text{Edd}$ \\
    & $[\text{erg}$ $\text{s}^\text{-1}$ $\text{Hz}^\text{-1}]$ & $[\text{erg}$ $\text{s}^\text{-1}$ $\text{Hz}^\text{-1}]$ & & $[\text{erg}$ $\text{s}^\text{-1}]$ & \\
  \hline
  \vspace*{+1mm} 
  0741390201 & 29.30 & 25.91 & 1.30 & ${45.40}^{+0.01}_{-0.02}$ & ${0.52}^{+0.02}_{-0.02}$ \\
  \vspace*{+1mm}
  0741390401 & 29.19 & 25.69 & 1.34 & ${45.15}^{+0.02}_{-0.03}$ & ${0.29}^{+0.02}_{-0.02}$ \\
  \hline
 \end{tabular}
\end{table*}

\subsection{Origin of the soft X-ray excess}
 
We investigated the \textit{XMM-Newton} X-ray band, which indicates a soft excess feature below 2 keV. In order to probe the origin of the soft X-ray excess, we tried the two most prevailing models, which are the relativistic blurred reflection from the ionized accretion disk and the thermal Comptonisation from a warm corona embedded in the surface of the disk. Despite the different physical scenarios, these two sets of models can both obtain a statistically good fit to the spectra. As for the relativistic reflection model, we tried three flavors of \texttt{relxill}, \texttt{relxillD} and \texttt{relxillCp} \citep{2014ApJ...782...76G}. The model \texttt{relxillD} only differs from \texttt{relxill} by allowing a higher density of the accretion disk. However, we found that the disk density is pegged at the lower limit of ${10}^{15}$ $\text{cm}^{-3}$, which is the fixed value of the density in \texttt{relxill}. Hence, there is literally no difference between these two flavors in our case, so it is not surprising that the ${\chi}^{2}$ did not even change for \texttt{relxillD} compared to \texttt{relxill}. The other model we tried is \texttt{relxillCp}, which presumed an incident spectrum by a nthcomp Comptonization continuum instead of the cutoff power law for \texttt{relxill}. Since the incident spectrum is physically produced by the thermal Comptonization of the seed photons from the disk in the hot corona, \texttt{relxillCp} is supposed to be a more accurate model for analyzing our source. Consequently, the fit quality of \texttt{relxillCp} proved it with a smaller ${\chi}^{2}$.However, when we look at the spectra and the fitted model, we notice an excess at the high energy end (above 5 keV). In the reflection spectra, there are always some reflection features, such as Fe K${\alpha}$ emission at around 6.4 keV. In our case, we did not find any Fe K${\alpha}$ features, except for a small peak between 6 and 7 keV in Obs. ID 0741390201 (Figure~\ref{fig:2}) which may be related to the iron emission. 

For the warm corona model, we chose \texttt{optxagnf} \citep{2012MNRAS.420.1848D} to analyze the spectra. This model is a combination of three components: (1) the thermal UV/optical emission from the accretion disk, (2) the thermal Comptonization of the UV/optical seed photons in a warm corona producing the soft X-ray excess, and (3) the seed photons from the disk getting Compton up-scattered in a hot corona to the primary power law continuum dominating the hard X-ray band. In the fitting procedure, we freezed the black hole spin to get an acceptable error for our fitted results. We used three distinctive values (0.0, 0.5, 0.998) to stand for non-spin, intermediate spin and maximal spin scenarios. Although these spins have totally different parameter values, especially for the bolometric luminosity, they all have similarly good fit statistics that it made hard to distinguish the right one for our source, thus we took into account the UV/optical data from SDSS, \textit{Swift} and GALEX to constrain the model. As shown in Figure~\ref{fig:4}, in the case of non-spin scenario, the extrapolated model to the UV/optical band can describe the data better than other cases. We find that the thermal Comptonization from a warm corona with non-spin can describe the spectra of SBS 1353+564 sufficiently.

\subsection{Variability}

The X-ray flux drops by a factor of $\sim$ 2 in just ten days between the two \textit{XMM-Newton} observations in 2014. As for this X-ray variability, there are two possible reasons: one is that the strength of Compton scattering decreases; the other may be attributed to the variation of the intrinsic UV photons. To find out the cause of the variability, we next consider the ability to transform UV photons to X-ray photons by calculating ${\alpha}_\text{\tiny OX}$ $=$ $-$ $\text{log}[L({\nu}_\text{\tiny X-ray})/L({\nu}_\text{\tiny UV})]/\text{log}({\nu}_\text{\tiny X-ray}/{\nu}_\text{\tiny UV})$ \citep{1979ApJ...234L...9T}. Following many previous works, we mainly focus on the luminosities at 2 keV and 2500 $\si{\angstrom}$, deriving the relationship of ${\alpha}_\text{\tiny OX}$ $=$ $-$ $0.3838$ $\text{log}({L}_\text{2 keV}/{L}_\text{\sevensize2500\si{\angstrom}})$ \citep{1979ApJ...234L...9T}. Table~\ref{tab:3} lists the UV and the X-ray properties, ${L}_\text{\sevensize2500\si{\angstrom}}$ and ${L}_\text{2 keV}$ are calculated from the fitted model under the non-spin scenario. The ${\alpha}_\text{\tiny OX}$ of the two observations are 1.30 and 1.34, which are not changing too much. The ${\alpha}_\text{\tiny OX}$ calculated here agree with the ${\alpha}_\text{\tiny OX}$-${L}_\text{\sevensize2500\si{\angstrom}}$ relation described as \citet[Equation (2)]{2006AJ....131.2826S} at the luminosity at 2500 $\si{\angstrom}$, and the ${\alpha}_\text{\tiny OX}$-${L}_\text{2 keV}$ relation described as \citet[Equation (3)]{2006AJ....131.2826S} at the luminosity at 2 keV. The ${\alpha}_\text{\tiny OX}$ of SBS 1353+564 is also given as 1.35 by \citet{2010ApJS..187...64G} using the \textit{Swift} data, which is quite close to our ${\alpha}_\text{\tiny OX}$ values. The two ${\alpha}_\text{\tiny OX}$ indicate that the ability to transform UV photons to X-ray photons is stable in a ten-day time. Hence, we guess that the variability of the X-ray spectra can be attributed to the variability of the intrinsic UV emission from the accretion disk. There are many mechanisms that can account for the UV variability. One possibility is the change of the accretion rate. We derive the bolometric luminosities (${L}_\text{bol}$) from the fitted parameter values in the non-spin scenario case (see Table~\ref{tab:2}). The Eddington ratios (${\lambda}_\text{Edd}$) of the two observations are ${0.52}^{+0.02}_{-0.02}$ and ${0.29}^{+0.02}_{-0.02}$. The corresponding bolometric luminosities are ${2.52}_{-0.09}^{+0.09}$ $\times$ $10^{45}$ $\text{erg}$ $\text{s}^{-1}$ and ${1.42}_{-0.08}^{+0.08}$ $\times$ $10^{45}$ $\text{erg}$ $\text{s}^{-1}$, respectively. The values are presented in Table~\ref{tab:3}. Other than the change of the accretion rate, another possibility to cause the UV variability can be due to the change of the magnetic field produced by the accretion disk \citep{1993ApJ...406..420M, 2009ApJ...698..895K, 2019MNRAS.483L..17D}. 

In summary, we have conducted for the first time the timing and the spectral analyses of the narrow-line Seyfert 1 galaxy, SBS 1353+564. We utilized the observations from \textit{XMM-Newton} and \textit{Swift}. We can conclude that:
\begin{itemize}
\item We find a prominent soft X-ray excess below 2 keV which cannot be well described by an added blackbody component.
\item We applied two physical models to describe the X-ray spectra. The relativistic blurred reflection model cannot fit the data above 5 keV and the spectra do not show any reflection features. Instead, the warm corona models with different spins derive different sets of best-fitting parameters.
\item By using the UV/optical data compared with the best-fit models extrapolated to the UV/optical band, we find that the warm corona model with non-spin scenario can describe the X-ray spectra of SBS 1353+564 well enough.
\end{itemize}
 
\section*{Acknowledgements}
\addcontentsline{toc}{section}{Acknowledgements}
We acknowledge the financial supports from the National Key R$\&$D Program of China grant 2017YFA0402703 and the National Natural Science Foundation of China grant 11733002. N.Ding acknowledges financial support from the scientific research fund of talent introduction of Kunming University and the support of scientific research fund of Yunnan Provincial Education Department (2021J0715). This work has made use of the data obtained from \textit{XMM-Newton} and \textit{swift} observations and the software {\sevensize HEASOFT} v.6.26.1, provided by NASA's High Energy Astrophysics Science Archive Research Center (HEASARC). We would like to thank those who have developed the \texttt{relxill} model.

\textit{Softwares}: SAS \citep{2004ASPC..314..759G}, XSPEC \citep{1996ASPC..101...17A}, Astropy \citep{2013A&A...558A..33A}, NumPy \citep{2011CSE....13b..22V}, Matplotlib \citep{2007CSE.....9...90H}, PyQSOFit \citep{2018ascl.soft09008G}.

\section*{Data availability}
\addcontentsline{toc}{section}{Data availability}
The data underlying this article will be shared on reasonable request to the corresponding author.

\bsp
\label{lastpage}

\begin{thebibliography}{99}
\bibitem[\protect\citeauthoryear{Arnaud}{1996}]{1996ASPC..101...17A} Arnaud K.~A., 1996, ASPC, 101, 17
\bibitem[\protect\citeauthoryear{Arnaud et al.}{1985}]{1985MNRAS.217..105A} Arnaud K.~A., Branduardi-Raymont G., Culhane J.~L., Fabian A.~C., Hazard C., McGlynn T.~A., Shafer R.~A., et al., 1985, MNRAS, 217, 105
\bibitem[\protect\citeauthoryear{Astropy Collaboration et al.}{2013}]{2013A&A...558A..33A} Astropy Collaboration, Robitaille T.~P., Tollerud E.~J., Greenfield P., Droettboom M., Bray E., Aldcroft T., et al., 2013, A\&A, 558, A33
\bibitem[\protect\citeauthoryear{Berton et al.}{2018}]{2018A&A...614A..87B} Berton M., Congiu E., J{\"a}rvel{\"a} E., Antonucci R., Kharb P., Lister M.~L., Tarchi A., et al., 2018, A\&A, 614, A87
\bibitem[\protect\citeauthoryear{Boller, Brandt, \& Fink}{1996}]{1996A&A...305...53B} Boller T., Brandt W.~N., Fink H., 1996, A\&A, 305, 53
\bibitem[\protect\citeauthoryear{Boller et al.}{2021}]{2021A&A...647A...6B} Boller T., Liu T., Weber P., Arcodia R., Dauser T., Wilms J., Nandra K., et al., 2021, A\&A, 647, A6
\bibitem[\protect\citeauthoryear{Brightman et al.}{2013}]{2013MNRAS.433.2485B} Brightman M., Silverman J.~D., Mainieri V., Ueda Y., Schramm M., Matsuoka K., Nagao T., et al., 2013, MNRAS, 433, 2485
\bibitem[\protect\citeauthoryear{Burrows et al.}{2005}]{2005SSRv..120..165B} Burrows D.~N., Hill J.~E., Nousek J.~A., Kennea J.~A., Wells A., Osborne J.~P., Abbey A.~F., et al., 2005, SSRv, 120, 165
\bibitem[\protect\citeauthoryear{Cheng et al.}{2020}]{2020MNRAS.495.1158C} Cheng H., Liu B.~F., Liu J., Liu Z., Qiao E., Yuan W., 2020, MNRAS, 495, 1158
\bibitem[\protect\citeauthoryear{Crummy et al.}{2006}]{2006MNRAS.365.1067C} Crummy J., Fabian A.~C., Gallo L., Ross R.~R., 2006, MNRAS, 365, 1067
\bibitem[\protect\citeauthoryear{Czerny et al.}{2003}]{2003A&A...412..317C} Czerny B., Niko{\l}ajuk M., R{\'o}{\.z}a{\'n}ska A., Dumont A.-M., Loska Z., Zycki P.~T., 2003, A\&A, 412, 317
\bibitem[\protect\citeauthoryear{Dexter \& Begelman}{2019}]{2019MNRAS.483L..17D} Dexter J., Begelman M.~C., 2019, MNRAS, 483, L17
\bibitem[\protect\citeauthoryear{Done et al.}{2012}]{2012MNRAS.420.1848D} Done C., Davis S.~W., Jin C., Blaes O., Ward M., 2012, MNRAS, 420, 1848
\bibitem[\protect\citeauthoryear{Dunn et al.}{2007}]{2007AJ....134.1061D} Dunn J.~P., Crenshaw D.~M., Kraemer S.~B., Gabel J.~R., 2007, AJ, 134, 1061
\bibitem[\protect\citeauthoryear{Edelson et al.}{2002}]{2002ApJ...568..610E} Edelson R., Turner T.~J., Pounds K., Vaughan S., Markowitz A., Marshall H., Dobbie P., et al., 2002, ApJ, 568, 610
\bibitem[\protect\citeauthoryear{Fitzpatrick}{1999}]{1999PASP..111...63F} Fitzpatrick E.~L., 1999, PASP, 111, 63
\bibitem[\protect\citeauthoryear{Gabriel et al.}{2004}]{2004ASPC..314..759G} Gabriel C., Denby M., Fyfe D.~J., Hoar J., Ibarra A., Ojero E., Osborne J., et al., 2004, ASPC, 314, 759
\bibitem[\protect\citeauthoryear{Garc{\'\i}a et al.}{2014}]{2014ApJ...782...76G} Garc{\'\i}a J., Dauser T., Lohfink A., Kallman T.~R., Steiner J.~F., McClintock J.~E., Brenneman L., et al., 2014, ApJ, 782, 76
\bibitem[\protect\citeauthoryear{Garc{\'\i}a et al.}{2013}]{2013ApJ...768..146G} Garc{\'\i}a J., Dauser T., Reynolds C.~S., Kallman T.~R., McClintock J.~E., Wilms J., Eikmann W., 2013, ApJ, 768, 146
\bibitem[\protect\citeauthoryear{George \& Fabian}{1991}]{1991MNRAS.249..352G} George I.~M., Fabian A.~C., 1991, MNRAS, 249, 352
\bibitem[\protect\citeauthoryear{Ghosh \& Laha}{2020}]{2020MNRAS.497.4213G} Ghosh R., Laha S., 2020, MNRAS, 497, 4213
\bibitem[\protect\citeauthoryear{Gierli{\'n}ski \& Done}{2004}]{2004MNRAS.349L...7G} Gierli{\'n}ski M., Done C., 2004, MNRAS, 349, L7
\bibitem[\protect\citeauthoryear{Goodrich}{1989}]{1989ApJ...342..224G} Goodrich R.~W., 1989, ApJ, 342, 224
\bibitem[\protect\citeauthoryear{Grupe}{2004}]{2004AJ....127.1799G} Grupe D., 2004, AJ, 127, 1799
\bibitem[\protect\citeauthoryear{Grupe et al.}{1999}]{1999A&A...350..805G} Grupe D., Beuermann K., Mannheim K., Thomas H.-C., 1999, A\&A, 350, 805
\bibitem[\protect\citeauthoryear{Grupe et al.}{1998}]{1998A&A...330...25G} Grupe D., Beuermann K., Thomas H.-C., Mannheim K., Fink H.~H., 1998, A\&A, 330, 25
\bibitem[\protect\citeauthoryear{Grupe et al.}{2010}]{2010ApJS..187...64G} Grupe D., Komossa S., Leighly K.~M., Page K.~L., 2010, ApJS, 187, 64
\bibitem[\protect\citeauthoryear{Grupe \& Mathur}{2004}]{2004ApJ...606L..41G} Grupe D., Mathur S., 2004, ApJL, 606, L41
\bibitem[\protect\citeauthoryear{Grupe et al.}{2004}]{2004AJ....127..156G} Grupe D., Wills B.~J., Leighly K.~M., Meusinger H., 2004, AJ, 127, 156
\bibitem[\protect\citeauthoryear{Guo, Shen, \& Wang}{2018}]{2018ascl.soft09008G} Guo H., Shen Y., Wang S., 2018, ascl.soft. ascl:1809.008
\bibitem[\protect\citeauthoryear{HI4PI Collaboration et al.}{2016}]{2016A&A...594A.116H} HI4PI Collaboration, Ben Bekhti N., Flöer L., et al., 2016, A\&A, 594, A116
\bibitem[\protect\citeauthoryear{Hill et al.}{2004}]{2004SPIE.5165..217H} Hill J.~E., Burrows D.~N., Nousek J.~A., Abbey A.~F., Ambrosi R.~M., Br{\"a}uninger H.~W., Burkert W., et al., 2004, SPIE, 5165, 217
\bibitem[\protect\citeauthoryear{Hunter}{2007}]{2007CSE.....9...90H} Hunter J.~D., 2007, CSE, 9, 90
\bibitem[\protect\citeauthoryear{Jansen et al.}{2001}]{2001A&A...365L...1J} Jansen F., Lumb D., Altieri B., Clavel J., Ehle M., Erd C., Gabriel C., et al., 2001, A\&A, 365, L1
\bibitem[\protect\citeauthoryear{Jefremov, Tsupko, \& Bisnovatyi-Kogan}{2015}]{2015PhRvD..91l4030J} Jefremov P.~I., Tsupko O.~Y., Bisnovatyi-Kogan G.~S., 2015, PhRvD, 91, 124030
\bibitem[\protect\citeauthoryear{Jiang et al.}{2020}]{2020MNRAS.498.3888J} Jiang J., Gallo L.~C., Fabian A.~C., Parker M.~L., Reynolds C.~S., 2020, MNRAS, 498, 3888
\bibitem[\protect\citeauthoryear{Kaspi et al.}{2000}]{2000ApJ...533..631K} Kaspi S., Smith P.~S., Netzer H., Maoz D., Jannuzi B.~T., Giveon U., 2000, ApJ, 533, 631
\bibitem[\protect\citeauthoryear{Kelly, Bechtold, \& Siemiginowska}{2009}]{2009ApJ...698..895K} Kelly B.~C., Bechtold J., Siemiginowska A., 2009, ApJ, 698, 895
\bibitem[\protect\citeauthoryear{Lu \& Yu}{1999}]{1999ApJ...526L...5L} Lu Y., Yu Q., 1999, ApJL, 526, L5
\bibitem[\protect\citeauthoryear{Lusso et al.}{2010}]{2010A&A...512A..34L} Lusso E., Comastri A., Vignali C., Zamorani G., Brusa M., Gilli R., Iwasawa K., et al., 2010, A\&A, 512, A34
\bibitem[\protect\citeauthoryear{Magdziarz et al.}{1998}]{1998MNRAS.301..179M} Magdziarz P., Blaes O.~M., Zdziarski A.~A., Johnson W.~N., Smith D.~A., 1998, MNRAS, 301, 179
\bibitem[\protect\citeauthoryear{Mangalam \& Wiita}{1993}]{1993ApJ...406..420M} Mangalam A.~V., Wiita P.~J., 1993, ApJ, 406, 420
\bibitem[\protect\citeauthoryear{Matzeu et al.}{2020}]{2020MNRAS.497.2352M} Matzeu G.~A., Nardini E., Parker M.~L., Reeves J.~N., Braito V., Porquet D., Middei R., et al., 2020, MNRAS, 497, 2352
\bibitem[\protect\citeauthoryear{Mundo et al.}{2020}]{2020MNRAS.496.2922M} Mundo S.~A., Kara E., Cackett E.~M., Fabian A.~C., Jiang J., Mushotzky R.~F., Parker M.~L., et al., 2020, MNRAS, 496, 2922
\bibitem[\protect\citeauthoryear{Nandra et al.}{2007}]{2007MNRAS.382..194N} Nandra K., O'Neill P.~M., George I.~M., Reeves J.~N., 2007, MNRAS, 382, 194
\bibitem[\protect\citeauthoryear{Nasa High Energy Astrophysics Science Archive Research Center (Heasarc)}{2014}]{2014ascl.soft08004N} Nasa High Energy Astrophysics Science Archive Research Center (Heasarc), 2014, ascl.soft. ascl:1408.004
\bibitem[\protect\citeauthoryear{Osterbrock \& Pogge}{1985}]{1985ApJ...297..166O} Osterbrock D.~E., Pogge R.~W., 1985, ApJ, 297, 166
\bibitem[\protect\citeauthoryear{Park et al.}{2006}]{2006ApJ...652..610P} Park T., Kashyap V.~L., Siemiginowska A., van Dyk D.~A., Zezas A., Heinke C., Wargelin B.~J., 2006, ApJ, 652, 610
\bibitem[\protect\citeauthoryear{Rakshit et al.}{2017}]{2017ApJS..229...39R} Rakshit S., Stalin C.~S., Chand H., Zhang X.-G., 2017, ApJS, 229, 39
\bibitem[\protect\citeauthoryear{Roming et al.}{2005}]{2005SSRv..120...95R} Roming P.~W.~A., Kennedy T.~E., Mason K.~O., Nousek J.~A., Ahr L., Bingham R.~E., Broos P.~S., et al., 2005, SSRv, 120, 95
\bibitem[\protect\citeauthoryear{Ross \& Fabian}{2005}]{2005MNRAS.358..211R} Ross R.~R., Fabian A.~C., 2005, MNRAS, 358, 211
\bibitem[\protect\citeauthoryear{Schlafly \& Finkbeiner}{2011}]{2011ApJ...737..103S} Schlafly E.~F., Finkbeiner D.~P., 2011, ApJ, 737, 103
\bibitem[\protect\citeauthoryear{Seibert et al.}{2012}]{2012AAS...21934001S} Seibert M., Wyder T., Neill J., Madore B., Bianchi L., Smith M., Shiao B., et al., 2012, AAS
\bibitem[\protect\citeauthoryear{Steffen et al.}{2006}]{2006AJ....131.2826S} Steffen A.~T., Strateva I., Brandt W.~N., Alexander D.~M., Koekemoer A.~M., Lehmer B.~D., Schneider D.~P., et al., 2006, AJ, 131, 2826
\bibitem[\protect\citeauthoryear{Str{\"u}der et al.}{2001}]{2001A&A...365L..18S} Str{\"u}der L., Briel U., Dennerl K., Hartmann R., Kendziorra E., Meidinger N., Pfeffermann E., et al., 2001, A\&A, 365, L18
\bibitem[\protect\citeauthoryear{Sunyaev \& Titarchuk}{1980}]{1980A&A....86..121S} Sunyaev R.~A., Titarchuk L.~G., 1980, A\&A, 500, 167
\bibitem[\protect\citeauthoryear{Tananbaum et al.}{1979}]{1979ApJ...234L...9T} Tananbaum H., Avni Y., Branduardi G., Elvis M., Fabbiano G., Feigelson E., Giacconi R., et al., 1979, ApJL, 234, L9
\bibitem[\protect\citeauthoryear{Thorne}{1974}]{1974ApJ...191..507T} Thorne K.~S., 1974, ApJ, 191, 507
\bibitem[\protect\citeauthoryear{Titarchuk}{1994}]{1994ApJ...434..570T} Titarchuk L., 1994, ApJ, 434, 570
\bibitem[\protect\citeauthoryear{Tomsick et al.}{2018}]{2018ApJ...855....3T} Tomsick J.~A., Parker M.~L., Garc{\'\i}a J.~A., Yamaoka K., Barret D., Chiu J.-L., Clavel M., et al., 2018, ApJ, 855, 3
\bibitem[\protect\citeauthoryear{Tripathi et al.}{2019}]{2019MNRAS.488.4831T} Tripathi S., Waddell S.~G.~H., Gallo L.~C., Welsh W.~F., Chiang C.-Y., 2019, MNRAS, 488, 4831
\bibitem[\protect\citeauthoryear{van der Walt, Colbert, \& Varoquaux}{2011}]{2011CSE....13b..22V} van der Walt S., Colbert S.~C., Varoquaux G., 2011, CSE, 13, 22
\bibitem[\protect\citeauthoryear{Vaughan et al.}{2003}]{2003MNRAS.345.1271V} Vaughan S., Edelson R., Warwick R.~S., Uttley P., 2003, MNRAS, 345, 1271
\bibitem[\protect\citeauthoryear{V{\'e}ron-Cetty \& V{\'e}ron}{2006}]{2006A&A...455..773V} V{\'e}ron-Cetty M.-P., V{\'e}ron P., 2006, A\&A, 455, 773
\bibitem[\protect\citeauthoryear{Vestergaard \& Peterson}{2006}]{2006ApJ...641..689V} Vestergaard M., Peterson B.~M., 2006, ApJ, 641, 689
\bibitem[\protect\citeauthoryear{Walton et al.}{2020}]{2020MNRAS.499.1480W} Walton D.~J., Alston W.~N., Kosec P., Fabian A.~C., Gallo L.~C., Garcia J.~A., Miller J.~M., et al., 2020, MNRAS, 499, 1480
\bibitem[\protect\citeauthoryear{Walton et al.}{2013}]{2013MNRAS.428.2901W} Walton D.~J., Nardini E., Fabian A.~C., Gallo L.~C., Reis R.~C., 2013, MNRAS, 428, 2901
\bibitem[\protect\citeauthoryear{Wilms, Allen, \& McCray}{2000}]{2000ApJ...542..914W} Wilms J., Allen A., McCray R., 2000, ApJ, 542, 914
\bibitem[\protect\citeauthoryear{Yang et al.}{2020}]{2020ApJ...904..200Y} Yang X., Yao S., Yang J., Ho L.~C., An T., Wang R., Baan W.~A., et al., 2020, ApJ, 904, 200
\bibitem[\protect\citeauthoryear{{\.Z}ycki, Done, \& Smith}{1999}]{1999MNRAS.309..561Z} {\.Z}ycki P.~T., Done C., Smith D.~A., 1999, MNRAS, 309, 561
\end{thebibliography}
\end{document}